\documentclass{aa}  

\usepackage{graphicx}
\usepackage{txfonts}
\usepackage{amsmath}
\usepackage{float}
\usepackage{ragged2e}
\usepackage{graphicx}
\usepackage{multirow} 
\usepackage{subcaption}
\usepackage{tablefootnote}
\usepackage{caption}
\usepackage[dvipsnames]{xcolor}
\usepackage{hyperref}

\graphicspath{{./}{figures/}}

\begin{document}

\title{X-ray properties of massive compact relic galaxies}

\author{
Orsolya E. Kov\'acs\inst{1,2},
Norbert Werner\inst{1},
\'Akos Bogd\'an \inst{2},
\and
Jelle de Plaa\inst{3}
}

\institute{
Department of Theoretical Physics and Astrophysics, Faculty of Science, Masaryk University, Kotl\'a\v{r}sk\'a 2, Brno, 611 37, Czech Republic, \email{o.e.kovacs@gmail.com}
\and
Center for Astrophysics\,\textbar\,Harvard \& Smithsonian, 60 Garden Street, Cambridge, MA 02138, USA
\and
SRON Netherlands Institute for Space Research, Niels Bohrweg 4, 2333 CA Leiden, The Netherlands
}

\date{Received Month XX, 2024; accepted Month YY, 2024}

\abstract
{We present the X-ray analysis of seven local compact elliptical galaxies (cEGs), selected for their morphological resemblance to high-redshift red nuggets. As likely descendants of the red nugget population, cEGs offer a unique window into the early Universe, enabling the study of early galaxy evolution and the interplay between black holes, stellar bulges, and dark matter halos.
Using data from \textit{Chandra} and \textit{XMM-Newton}, we investigate the properties of the hot gaseous halos in cEGs. Two galaxies -- MRK\,1216 and PGC\,32873 -- host luminous, spatially extended X-ray atmospheres, allowing us to derive radial thermodynamic profiles.
For MRK\,1216, we performed high-resolution spectral modeling with RGS data, which hints at super-solar $\alpha/\rm{Fe}$ abundance ratios.
The remaining galaxies show either faint or undetected X-ray halos, though several display AGN-like (active galactic nucleus) power-law emission.
In the context of local scaling relations, cEGs show only mild deviations from the general galaxy population, exhibiting a slightly steeper $M_{\star}-L_{X}$ relation and occupying the lower boundary of the $M_{\star}$–$M_{\rm vir}$ relation.
These trends suggest that high-redshift red nuggets may also host a diverse range of X-ray atmospheres. We speculate that the compactness of cEGs may trace back to the population of `little red dots' (LRDs), hinting at a potential link between LRDs, red nuggets, and compact relic galaxies in the local Universe.}

\keywords{
X-rays: galaxies --
galaxies: elliptical and lenticular, cD --
galaxies: evolution --
galaxies: formation
}

\titlerunning {X-ray properties of massive compact relic galaxies}
\authorrunning {Kov\'acs et al.}

\maketitle

\begin{table*}
\caption{{Galaxy properties}}
\label{tab:gal-prop}
\centering
\begin{tabular}{lccrccrcc} 
\hline\hline        
\multicolumn{1}{c}{\multirow{2}{*}{Galaxy}} &  \multirow{2}{*}{R.A.} & \multirow{2}{*}{Decl.} & \multicolumn{1}{c}{D} & $\mathrm{log}\,M_{*}$ & \multicolumn{1}{c}{$r_{\mathrm{eff}}$} & \multicolumn{1}{c}{$\sigma$} & \multicolumn{1}{c}{$N_{\mathrm H}$} &\multicolumn{1}{c}{$N_{\mathrm{gal}}$}\\ 
&&& \multicolumn{1}{c}{[Mpc]} & $[M_{\odot}]$ & \multicolumn{1}{c}{[kpc]} & \multicolumn{1}{c}{[$\rm km \, s^{-1}$]} & \multicolumn{1}{c}{[$10^{20} \, \rm cm^{-2}$]} & \multicolumn{1}{c}{(<0.2\,\rm Mpc)}\\
\hline
MRK\,1216$^{\dagger}$ & 08 28 47.11 & $-$06 56 24.44 & $94 \pm 2$ & $11.34^{+0.11}_{-0.10}$ & $3.0 \pm 0.1$ & $308 \pm 7$ & 4.58 & 0\\
NGC\,472$^{\dagger}$ & 01 20 28.70 & +32 42 32.68 & $74 \pm 1$ & $11.07^{+0.06}_{-0.11}$ & $2.4 \pm 0.1$ & $217\pm8$ & 6.52 & 1\\
PGC\,32873$^{\dagger}$ & 10 56 16.06 & +42 19 59.30 & $112 \pm 2$ & $11.28^{+0.04}_{-0.04}$ & $2.3 \pm 0.1$ & $304\pm8$ & 1.21 & 1\\
PGC\,70520$^{\dagger}$ & 23 07 20.19 & +36 21 45.14 & $72 \pm 1$ & $10.95^{+0.10}_{-0.12}$ & $1.6 \pm 0.1$ & $248\pm8$ & 10.30 & 0\\
J091842.83+031557.1$^{\star}$ & 9 18 42.83 & +03 15 57.10 & 239.5 & $10.60^{+0.04}_{-0.05}$ & 1.2 & $221 \pm 8$ & 3.16 & 0\\
J092318.89+025538.3$^{\star}$ & 09 23 18.88 & +02 55 38.30 & 395.5 & $10.63^{+0.07}_{-0.11}$ & 1.8 & $227\pm12$ & 3.92 & 0\\
J093634.60+034428.4$^{\star}$ & 09 36 34.59 & +03 44 28.40 & 202.6 & $10.97^{+0.04}_{-0.05}$ & 2.1 & $256\pm5$ & 3.13 & 0\\
\hline 
\end{tabular}
\tablefoot{$^{\dagger}\,$The $D$ distances, $M_*$ stellar masses, $r_{\mathrm{eff}}$ effective radii, and $\sigma$  central stellar velocity dispersions are adopted from \cite{2017MNRAS.468.4216Y}; $M_*$ is obtained from dynamical models, $r_{\mathrm{eff}}$ is measured along the major axis of the best-fitting ellipse containing half of the light, and $\sigma$ is given within the best-fitting circular aperture. $^{\star}\,$$D$, $M_*$, and $\sigma$ are adopted from SDSS spectroscopic catalogs.
$N_{\mathrm{H}}$ is the weighted H\,I+H\,II galactic hydrogen column density along the given sightline \citep{2013MNRAS.431..394W}. {$N_{\mathrm{gal}}$ shows the number of neighboring galaxies within $0.2$\,Mpc distance determined from a cone search}.}
\end{table*}

\section{Introduction} \label{sec:introduction}
According to cosmological simulations, the core of massive, early-type galaxies forms through a rapid dissipative collapse ($z > 2$), with subsequent evolution to massive early-type galaxies through mergers ($z > 0$) \citep{2010ApJ...725.2312O}. Due to the stochasticity of the second phase, a small fraction of the collapsed cores could avoid mergers, leading to the formation of red nuggets observed at $1<z<2$ \citep{2009ApJ...695..101D}. During their further evolution, red nuggets may continue their undisturbed evolution, resulting in a rare population of massive, compact relic galaxies \citep{2013ApJ...773L...8Q} with only a small number of confirmed cases in the local universe \citep{2014ApJ...780L..20T,2017MNRAS.467.1929F}. The estimated fraction of massive galaxies that formed at $z > 2$ and survived structurally intact to $z \sim 0.$ is $<\!2\%$ \citep{2013ApJ...773L...8Q,2015MNRAS.449..361W}.

Local relic galaxies are outliers in the mass$-$size relation of local ellipticals: despite their massive nature ($M_{\star} \gtrsim 10^{11} M_{\odot}$) they are extremely compact ($r_{\mathrm{eff}} \lesssim 2$\,kpc), and host a very old ($\gtrsim\!13$\,Gyr) and passive (red) stellar population.
Their red and compact nature also resembles the `little red dot' (LRD) galaxy population, which is abundant in the $z \geq 4$ Universe; however, unlike typical LRDs, these galaxies do not exhibit the peculiar V-shaped spectral energy distribution (SED) characteristics of that population.

Local relics may be more complex than their optical appearance suggests, and exploring them across other wavelengths could offer a deeper understanding of their nature and origin. X-ray observations, in particular, provide valuable insights into their large-scale dark matter halos. Indeed, \textit{Chandra} observations revealed that two compact relic galaxies, MRK\,1216 and PGC\,32873, host extended, luminous X-ray atmospheres \citep{2018ApJ...854..143B,2018MNRAS.477.3886W,2019ApJ...877...91B}, implying that these galaxies reside in massive dark matter halos.
Although both sources are isolated, they exhibit very different X-ray properties \citep{2018MNRAS.477.3886W}, revealing that the population of local relics is less homogeneous than their optical appearance suggests.

{MRK\,1216 and PGC\,32873 have also been proposed to host over-massive black holes at their center \citep{2015ApJ...808...79F,2018MNRAS.477.3886W}. In fact, \cite{2015ApJ...808...79F} suggested that massive relic galaxies, by their very nature, host over-massive black holes: After the dissipative collapse of the core and the formation of the central supermassive black hole (BH), the galaxy does not go further in mass or size, thereby preventing the stellar bulge from catching up to its initially over-massive BH. In addition, X-ray studies show that nearby galaxies with over-massive BHs, such as NGC\,1332 and NGC\,4342, tend to reside in extended dark matter (DM) halos \citep{2012ApJ...753..140B,2018MNRAS.477.3886W,2019ApJ...877...91B}. Given the strong correlation between DM halo mass and BH mass, these findings suggested that BH growth is governed by the large-scale DM halo, and challenged the bulge–BH co-evolution \citep[e.g.][]{2024NatAs...8..126B,2024ApJ...965L..21K}. The presence of extended DM halos also implies that the compact nature of MRK\,1216 and PGC\,32873 is not a result of tidal stripping, as the DM halo would be stripped before the stellar component due to its low specific binding energy \citep{2011MNRAS.411.1525L}.}

{Overall, local relic galaxies, with their growth ceased at $z\sim2$, offer a glimpse into the early Universe and a unique opportunity to study early galaxy evolution and the interplay between different galactic components: the BH, the stellar bulge, and the DM halo. In this paper, we analyze \textit{Chandra} and \textit{XMM-Newton} observations to investigate the X-ray atmospheres of seven compact elliptical galaxies (cEGs) in the local Universe, including two confirmed local relics, Mrk\,1216 and PGC\,32873. We infer total masses from X-ray luminosities, estimate black hole masses from velocity dispersions, and adopt stellar masses from \textit{HST} data, placing these in the context of local scaling relations. In addition, we derive the radial profiles of thermodynamic properties for both MRK\,1216 and PGC\,32873, and perform spectral modeling of the high-resolution RGS data for MRK\,1216.}

This paper is structured as follows. Section\,\ref{sec:sample} introduces the galaxy sample. Section\,\ref{sec:obs} outlines the data reduction process. Section\,\ref{sec:res} presents the analysis and results, including point source identification (Section\,\ref{sec:res-ps}), source characterization (Section\,\ref{sec:res-hr}), analysis of diffuse emission (Sections\,\ref{sec:res-de}), surface brightness and thermodynamic profiles (Sections\,\ref{sec:sb-prof} and \ref{sec:thermo-prof}), and the high-resolution spectrum of MRK\,1216 (Section\ref{sec:hi-res-mrk}). A discussion is presented in Section\,\ref{sec:discussion}, including a comparison with the general galaxy population (Section\,\ref{sec:discussion1}) and the connection to little red dots (Section\,\ref{sec:discussion1}). A summary of the findings are given in Section\,\ref{sec:summary}.

We adopt a standard $\Lambda$CDM cosmology throughout the analysis with $H_{0}\,=\,71\, \rm km^{-1} \, s^{-1} \,Mpc^{-1} $, $\rm\Omega_M = 0.27$, and $\rm\Omega_{\Lambda} = 0.73$. 

\section{Sample} \label{sec:sample}
In this work, we study seven nearby cE galaxies. To construct a comprehensive sample, our targets were drawn from two distinct samples, with each subset exhibiting the typical characteristics of the relic population.

A subset of 4 galaxies is derived from the optically selected sample of \cite{2017MNRAS.468.4216Y}. This parent sample features 16 local cEs with detailed Hubble Space Telescope (HST) based measurements of their stellar mass, size, and structural and dynamical properties. 
{To select our sample, we first investigated the environment of the galaxies. Interestingly, simulations \citep{2016MNRAS.461..156P} and observations \citep{2018A&A...619A.137B} suggest that low-redshift massive relic galaxies are more common in rich environments (e.g. NGC1277 \citep{2014ApJ...780L..20T}, the prototypical relic galaxy near the center of Perseus cluster) than in isolation. As early members of their clusters, they tend to be centrally located with high relative velocities, which lowers their merger rates and allows them to remain intact.}
However, tidally stripped galaxies, with otherwise conventional evolutionary paths, can mimic some of the main optical characteristics  of low-redshift relics. Stripping and merging events are more common in dense cluster environments, where the level of X-ray emission from the intra-cluster medium may surpass that of individual galaxies, hampering the study of their X-ray atmospheres.
Therefore, from the parent sample, we selected only those galaxies that reside in low-density environments, as these provide a cleaner and less contaminated sample compared to those in rich environments. Environmental features (cluster and group environment, or interaction with another galaxy) narrowed the parent sample down to four galaxies: MRK\,1216, NGC\,422, PGC\,32873, and PGC\,70520 (Table\,\ref{tab:gal-prop}). Among these, MRK\,1216 and PGC\,32873 have already been the subjects of previous X-ray studies \citep{2018ApJ...854..143B,2018MNRAS.477.3886W}; however, additional observations acquired since then allow for a more in-depth analysis.

We extended the sample by incorporating $3$ additional X-ray luminous {cE} galaxies at $z<0.1$ (J091842.83+031557.1, J092318.89+025538.3, and J093634.60+034428.4) from the $140\, \mathrm{deg}^2$ \textit{eROSITA} Final Equatorial Depth Survey (eFEDS) field that have relatively high velocity dispersions and stellar masses (Table\,\ref{tab:gal-prop}). This extension aims to build a more comprehensive sample of X-ray-detected cE galaxies and may serve as a pathfinder for future studies targeting X-ray-selected compact systems with extended hot halos. Note that these three galaxies lie beyond the 112\,Mpc distance limit of the \cite{2017MNRAS.468.4216Y} sample and were therefore not included in their selection.
Based on SDSS data, the morphology and color indices ($u - r = 2.82 - 3.00$) indicate that these are early-type galaxies. Their effective radii are $\sim50\%$ smaller than expected for their mass at low redshift, but align with those of $z\sim2$ galaxies. A cone search and cross-correlation with the \textit{eROSITA} eFEDS cluster and group catalog \citep{2022A&A...661A...2L} confirm that the target galaxies are isolated sources, indicating passive evolution. Together, these characteristics suggest their classification as red nugget analogs.
The presence of a luminous X-ray halo in these galaxies was suggested by the \textit{eROSITA} X-ray observations. As a caveat, we note that eROSITA's broad point spread function (PSF) limits resolution.

Thus, the complete sample consists of 7 galaxies; their properties and X-ray observational log are detailed in Table\,\ref{tab:gal-prop} and Table\,\ref{tab:obs-log}, respectively. {Table\,\ref{tab:gal-prop}  also lists the number of neighboring galaxies within $200$\,kpc based on the \texttt{Radial Velocity Constrained Cone Search} tool in NED. While NGC\,472 and PGC\,32873 each have one such neighbor, their stellar masses -- estimated based on NIR magnitudes from the formula of \cite{2003ApJS..149..289B} -- are about three orders of magnitude lower than that of the target galaxy. The remaining galaxies have no companions within the search radius. This indicates that our targets lack close, massive companions.}
\textcolor{black}{For the most massive systems, MRK\,1216 and PGC\,32873, we repeated the cone search with an increased search radius of 1\,Mpc, yielding a total of three and five neighboring galaxies, respectively. These companions span stellar masses from the dwarf-galaxy regime ($\lesssim 10^8\,M_{\odot}$) to intermediate-mass galaxies ($\sim10^{10}\,M_{\odot}$), i.e. at least an order of magnitude lower than the central galaxies.}

\section{Data reduction} \label{sec:obs}

We carried out X-ray analysis on a combination of \textit{Chandra} ACIS-S, \textit{Chandra} HRC-I, and \textit{XMM-Newton} observations incorporating all available X-ray data of the targets (Table\,\ref{tab:obs-log}). After retrieving the data, we performed all subsequent processing with the CIAO software packages (Chandra Interactive Analysis of Observations, version 4.13, \cite{2006SPIE.6270E..1VF}, CALDB version 4.9.5) and  the XMM-SAS software packages (Science Analysis System, version 20.0.0, 
\cite{2017xru..conf...84G}).
We followed the standard data processing described, e.g., in \cite{2023A&A...678A..91K} and \cite{2017ApJ...850...98B} to prepare the X-ray observations for analysis.
{Specifically, for \textit{Chandra}, we reprocessed the data using \texttt{chandra\_repro} to ensure consistent calibration corrections across all observations, followed by the identification and removal of flare-contaminated intervals. For EPIC data, we additionally filtered out bad pixels, bad columns, cosmic rays, and events outside the FOV.}
From the reprocessed and cleaned data, we extracted energy-filtered event files and images, as well as exposure maps to correct for vignetting and variations of quantum efficiency across the CCD. 
{For the imaging analysis, the EPIC data were processed separately for each observation and detector, while the \textit{Chandra} data were merged from multiple observations. Spectra were extracted from individual observations and the resulting spectra and response files were subsequently combined for ACIS data.}

{The \textit{XMM-Newton} RGS data for MRK 1216 were extracted following the data processing steps described in \citet{deplaa2017}. The RGS spectrum has been extracted from a 0.5 arcmin-wide region in cross-dispersion direction to probe the X-ray brightest central part of the system. Because the RGS operates without a slit, every line is broadened by the spatial extent of the source. To account for this broadening in our model, we convolved the RGS response with the surface brightness profile of the galaxy derived from the EPIC/MOS1 image in the $0.5-1.8$~keV ($7-25$~\AA) band along the dispersion direction. However, because of the possible presence of intrinsic metallicity gradients, this surface brightness profile is not necessarily a perfect description of the observed line profile.  Therefore, in our model, we multiplied the line profile by a scale factor $s$, which is the ratio of the observed line spread function (LSF) width to the expected LSF. We leave this scale factor as a free parameter in our fits.}  

{Throughout the analysis, background subtraction was carried out using a local background region extracted from the same chip as the source emission.}

\begin{table}
\caption{Observational log}
\label{tab:obs-log}
\centering
\addtolength{\tabcolsep}{0.15em}
\begin{tabular}{lcccr}
\hline\hline        
\multicolumn{5}{c}{\multirow{2}{*}{\textit{Chandra}}} \\
&&&&\\
\hline
\multicolumn{1}{c}{\multirow{2}{*}{Galaxy}} & \multirow{2}{*}{Obs. ID} & \multirow{2}{*}{Obs. date} & \multirow{2}{*}{Instr.} & \multicolumn{1}{c}{$t_{\mathrm{exp}}$}\\ 
&&&& \multicolumn{1}{c}{[ks]} \\
\hline
MRK\,1216 & 17061 & 2015-06-12 & ACIS-S & 12.89\\
 & 20342 & 2018-01-09 & ACIS-S & 31.65\\
 & 20924 & 2018-01-09 & ACIS-S & 29.67\\
 & 20925 & 2018-01-12 & ACIS-S & 31.65\\
 & 20926 & 2018-01-14 & ACIS-S & 29.68\\
NGC\,472 & 25227 & 2023-07-25 & HRC-I & 9.75\\
 & 27962 & 2023-07-27 & HRC-I & 9.93 \\
 & 27963 & 2023-07-27 & HRC-I & 14.07\\
 & 25864 & 2023-08-01 & HRC-I & 14.07\\
 & 27973 & 2023-08-02 & HRC-I & 14.09\\
PGC\,32873 & 17063 & 2015-03-02 & ACIS-S & 23.38\\
 & 21377 & 2019-01-22 & ACIS-S & 57.31\\
 & 22061 & 2019-01-26 & ACIS-S & 31.65\\
 & 22062 & 2019-01-27 & ACIS-S & 21.78\\
 & 21378 & 2019-02-11 & ACIS-S & 24.74\\
 & 22100 & 2019-02-12 & ACIS-S & 24.74\\
 & 22101 & 2019-02-13 & ACIS-S & 33.62\\
 & 22102 & 2019-02-11 & ACIS-S & 23.26\\
PGC\,70520 & 25226 & 2023-08-10 & HRC-I & 13.77\\
 & 28368 & 2023-08-10 & HRC-I & 12.83\\
 & 27490 & 2023-08-15 & HRC-I & 14.28\\
 & 28485 & 2023-08-17 & HRC-I & 14.25\\
\end{tabular}
\addtolength{\tabcolsep}{-0.25em}
\begin{tabular}{lccr} 
\hline\hline
\multicolumn{4}{c}{\multirow{2}{*}{\textit{XMM-Newton}}} \\
&&&\\
\hline
\multicolumn{1}{c}{\multirow{2}{*}{Galaxy}} & \multirow{2}{*}{Obs. ID} & \multirow{2}{*}{Obs. date} & \multicolumn{1}{c}{$t_{\mathrm{exp}}$}\\ 
&&& \multicolumn{1}{c}{[ks]} \\
\hline
MRK\,1216 & 0822960101 & 2018-05-07 &  140.8\\
& 0822960201 & 2018-11-01 &  109.7\\
NGC\,472 & 0892400201 & 2022-02-09 & 40.9\\
PGC\,32873 & 0842730101 & 2020-04-26 & 87.0\\
& 0842730201 & 2020-05-11 & 83.0\\
PGC\,70520 & 0892400101 & 2022-12-05 & 45.9\\
J091842.83+031557.1 & 0922260301 & 2024-04-21 &  65.4 \\
J092318.89+025538.3 & 0922260201 & 2023-11-12 &  35.7 \\
J093634.60+034428.4 & 0922260101 & 2024-05-04 &  17.3 \\
\hline 
\end{tabular}
\end{table}

\section{Analysis and results} \label{sec:res}

\begin{table*}
\caption{{Unabsorbed X-ray luminosities of the galactic diffuse emission measured within $20\,r_{\mathrm{eff}}$}}
\label{tab:measured}
\centering
\begin{tabular}{lcccc} 
\hline\hline        
\multicolumn{1}{c}{\multirow{2}{*}{Galaxy}} & \multicolumn{1}{c}{$L^{\mathrm{ACIS-S}}_{0.5-2\,\mathrm{keV}}$} & $L^{\mathrm{MOS1}}_{0.5-2\,\mathrm{keV}}$ & $L^{\mathrm{MOS2}}_{0.5-2\,\mathrm{keV}}$ & $L^{\mathrm{PN}}_{0.5-2\,\mathrm{keV}}$\\ 
& \multicolumn{1}{c}{$[\rm erg \, s^{-1}]$} & $[\rm erg \, s^{-1}]$ & $[\rm erg \, s^{-1}]$\\
\hline
MRK\,1216   & $(9.16\pm0.13)\times 10^{41}$ & \multicolumn{2}{c}{$(8.14\pm0.06)\times10^{41}$}  & $(7.16\pm0.04)\times10^{41}$\\
NGC\,472    & $\hspace{0.9cm}<4.47\times10^{40}$        & $(2.33\pm0.69)\times10^{40}$                      &  $\hspace{0.9cm}<1.21\times10^{40}$ & $(4.06\pm1.35)\times10^{39}$\\
PGC\,32873  & $(1.09\pm0.06)\times10^{41}$  & \multicolumn{2}{c}{$(9.18\pm0.23)\times10^{40}$}  & $(7.33\pm0.16)\times10^{40}$\\
PGC\,70520  & $\hspace{0.9cm}<2.70\times10^{40}$        & $(4.88\pm1.92)\times10^{39}$ & $(3.39\pm1.75)\times10^{39}$  & $(4.00\pm1.04)\times10^{39}$\\
\hline 
\end{tabular}
\newline
\tablefoot{Luminosities are determined by spectral modeling for MRK\,1216 and PGC\,32873, and by photometric measurements for NGC\,472 and PGC\,70520. The spectra from MOS1 and MOS2 were fitted simultaneously. Error bars represent $1\,\sigma$ statistical uncertainties assuming Gaussian statistics; for non-detections, we present the $1\,\sigma$ upper limits. {The eFEDs targets are not listed, as their \textit{XMM-Newton} data is dominated by a central AGN, with no significant diffuse emission detected (Section \ref{sec:res-hr}).}}
\end{table*}

\begin{table*}[]
\caption{Best-fit spectral model parameters within $10\,r_{\rm{eff}}$}
\label{tab:spec-fit}
\centering
\begin{tabular}{ll|ccc|ccc}
\hline\hline
\multirow{2}{*}{Component}            & \multirow{2}{*}{Parameter}  & \multicolumn{3}{c}{Mrk\,1216} & \multicolumn{3}{c}{PGC\,32873}\\
\cline{3-8}
                           &                  & ACIS      & MOS      & PN     & ACIS      & MOS      & PN      \\
\hline
\multirow{2}{*}{apec}      & $kT$ [keV]       & $0.80\pm0.05$ & $0.76\pm0.00$ & $0.75\pm0.00$ & $0.71\pm0.03$ & $0.79\pm0.01$ & $0.76\pm0.01$ \\
                           & Abund.        & $0.37\pm0.04$ & $0.37$* & $0.37$* & $0.22\pm0.08$ & $0.22$* & $0.22$* \\
\hline
\multirow{1}{*}{apec}      & $kT$ [keV]       & $0.53\pm0.12$ & -- & -- & -- & -- & -- \\
\hline
\end{tabular}
\begin{flushleft}
\footnotesize{$^*$ Abundances are fixed to the best-fit \textit{Chandra} abundance.}
\end{flushleft}
\end{table*}

\subsection{Point sources} \label{sec:res-ps}
For galaxies with \textit{XMM-Newton} and \textit{Chandra} HRC data (NGC\,472 and PGC\,70520), we primarily used the HRC observations to identify X-ray point sources, taking advantage of its superior spatial resolution ($\sim\!0.4\,\arcsec$) compared to \textit{XMM-Newton}.
We ran the CIAO \texttt{wavdetect} source detection tool on the wide-band {($\sim0.1-10$\,keV)}\footnote{{Note that the HRC camera has no useful energy resolution.}} total counts mosaic of each galaxy, specifying $6$ wavelet scales {with values of 1, 1.414, 2, 2.828, 4, 8, and 16,} and a signal threshold of $<10^{-5}$. A single point source was identified near the center of PGC\,70520 with an offset of $0.27\,\arcsec$.
No further point sources were detected within $5\,r_{\rm eff}$ projected distance from these galaxies.

The flux of the point source was determined with the \texttt{srcflux} CIAO tool extracted within a circular region enclosing 90\% of the PSF at 1.0\,keV with the source's PSF fraction estimated from MARX simulations \citep{2012SPIE.8443E..1AD}. We applied an absorbed power-law model for the source spectrum with a photon index of $\Gamma = 1.7$, then converted to luminosity assuming that the source is associated with PGC\,70520. We measured an HRC wide-band source luminosity of $L_{\mathrm{wide}} = (1.17^{+0.48}_{-0.45})\times 10^{40} \, \rm erg \, s^{-1}$ on the merged data. Given this luminosity, the source emission {most likely originates} from a {low-luminosity} AGN, since LMXB luminosity functions typically cut off above $2\times10^{39}\,\rm erg \, s^{-1}$ \citep{2003ApJ...587..356I,2004MNRAS.349..146G,2004ApJ...611..846K,2012A&A...546A..36Z}. 

We ran \texttt{wavdetect} on the merged broad-band ACIS images of those galaxies with \textit{Chandra} ACIS data, MRK\,1216 and PGC\,32873, and identified $1$ and $2$ point sources, respectively, within $5\,r_{\rm eff}$ projected distances. These sources are likely background AGN, as none of them are within $2\,r_{\rm eff}$, where most of a galaxy's optical light is concentrated. Note that the detection sensitivity is relatively high due to the presence of bright diffuse emission in these galaxies.

To identify point sources in the EPIC images, we ran \texttt{wavdetect} on individual observations for each camera in the $0.5$–$10$\,keV band, using a significance threshold of $<10^{-4}$ and wavelet scales following the "$\sqrt{2}$ sequence", i.e., 1, $\sqrt{2}$, 2, $2\sqrt{2}$, 4, $4\sqrt{2}$, 8, $8\sqrt{2}$, 16. We identified a point-like source at the coordinates of each galaxy.

\subsection{Characterization of eFEDS galaxies observed only with \textit{XMM Newton}} \label{sec:res-hr}
All three galaxies in our sample with only \textit{XMM-Newton} EPIC data (J091842.83+031557.1, J092318.89+025538.3, and J093634.60+034428.4) show detectable X-ray emission, which is consistent with a point-like source. However, due to the broader XMM-Newton PSF, we performed additional tests to assess whether these sources are point-like or extended.

To infer the nature of the emission, we first compared the sources' radial brightness profiles with the PSF using the \texttt{eradial} tool. We found that the $0.5-7$\,keV emission from all three galaxies is consistent with the nominal PSF, with no significant X-ray excess, suggesting that the sources are point-like.

We further investigated the emission by analyzing its spectral properties. For this, we calculated the galaxies' hardness ratios, as spectral analysis was not feasible due to the faint nature of the sources (the brightest source has $\lesssim45$ net counts in the $0.5-7$\,keV band within $5\,r_{\mathrm{eff}}$). We used the BEHR code \citep{2006ApJ...652..610P}, and calculated the hardness ratio for each galaxy as a fractional difference defined by $(H-S)/(H+S)$, where $H$ and $S$ are the source counts in the hard ($2-10$\,keV) and soft ($0.3-2$\,keV) bands. For reference, we performed the same calculation for simulated power-law and APEC spectra, each tailored to the observations corresponding to the galaxies. We used \texttt{Xspec}'s\,\citep{1996ASPC..101...17A} \texttt{fakeit} command along with the appropriate response files, assuming $\Gamma = 1.7$ for the power-law model, and $kT=0.3$\,keV and an abundance of $0.3$ for the APEC model to perform the simulations. From the comparison of observed and simulated hardness ratios, we determined that all three sources exhibit predominantly power-law emission, suggesting that they are AGN. 

The X-ray emission of these galaxies is likely dominated by a central AGN. Studying their CGM requires separating this point source contribution. However, due to the broad on-axis PSF of \textit{XMM-Newton}’s EPIC cameras ($\sim6\,\arcsec$), which is comparable to the CGM extent observed in MRK\,1216 and PGC\,32873, the diffuse emission cannot be resolved in these systems.

\begin{figure*}[hpt]
\centering
\resizebox{0.46\hsize}{!}{\includegraphics{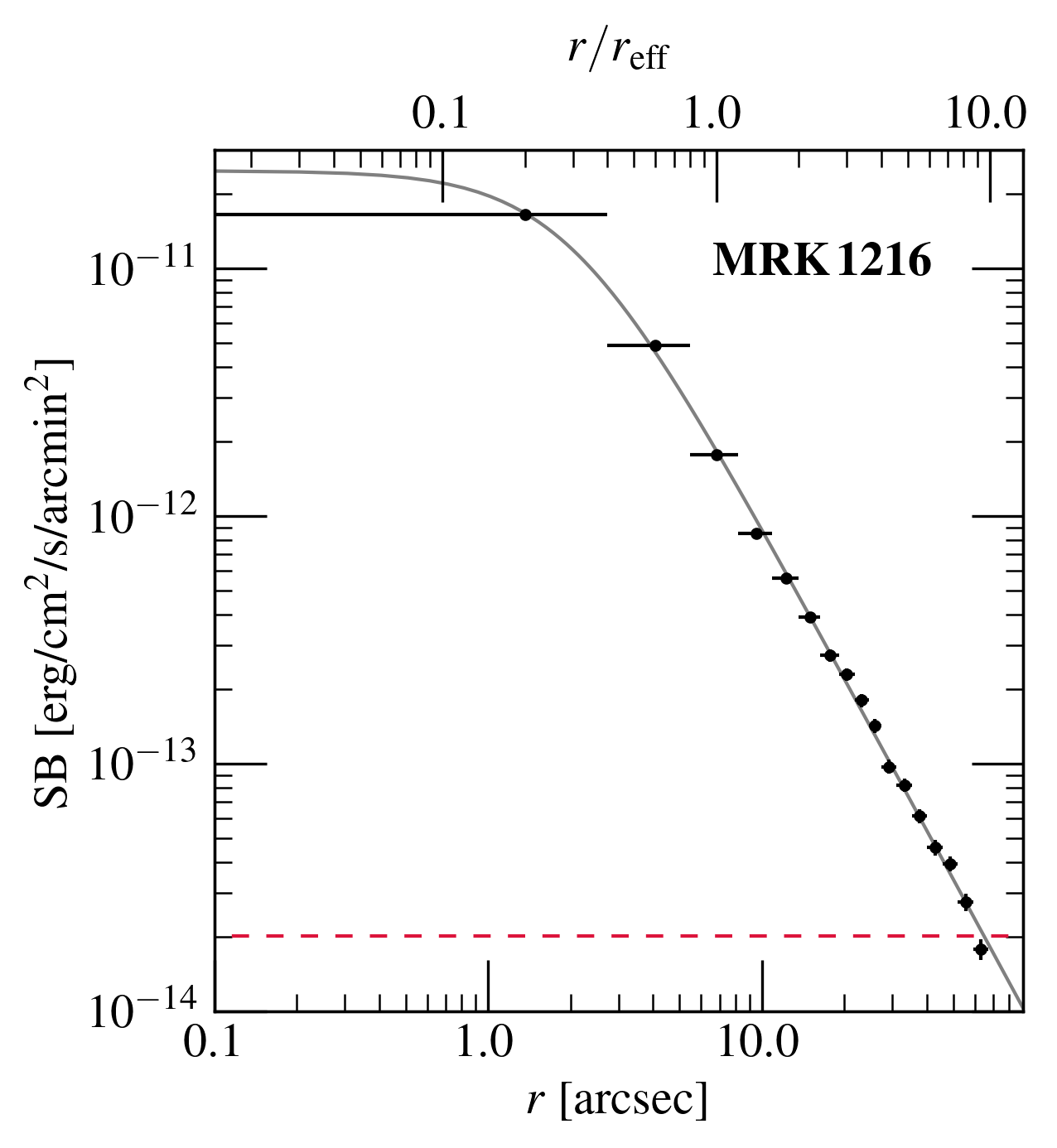}}
\hspace{0.05\textwidth}
\resizebox{0.46\hsize}{!}{\includegraphics{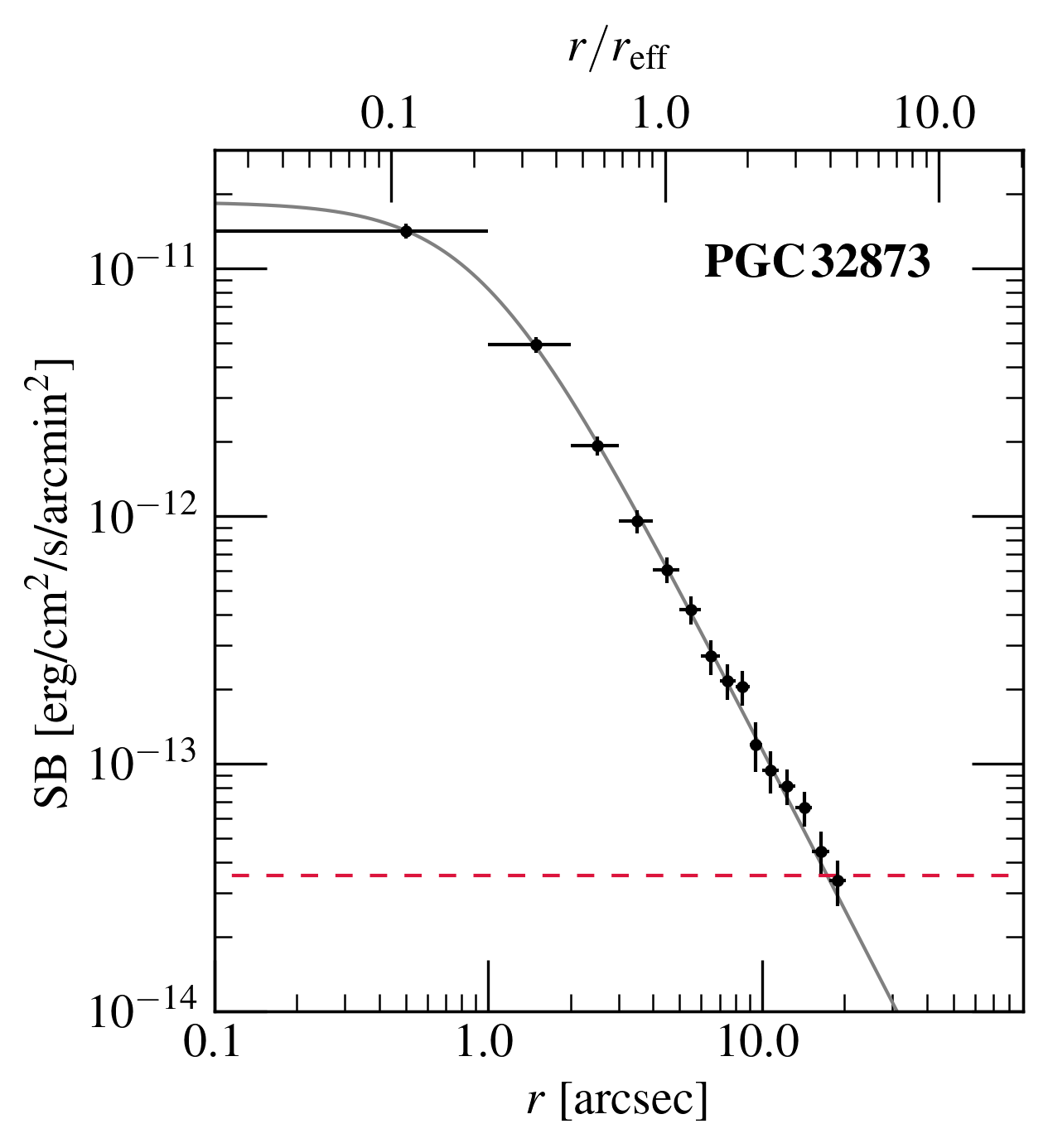}}
\caption{The $0.3-2$\,keV surface brightness profile of MRK\,1216 (left panel) and PGC\,32873 (right panel), and the best-fitting $\beta$-model with $\beta=0.507\pm0.003$, $r_{\mathrm{c}} = 0.874\pm0.001$\,kpc and $\beta=0.526\pm0.011$, $r_{\mathrm{c}}=0.492\pm0.001$\,kpc, respectively.}
\label{fig:sbp}
\end{figure*}

\subsection{Diffuse emission} \label{sec:res-de}

To characterize the galaxies' diffuse emission, we relied on ACIS and EPIC observations. 
For apertures with over $200$ net counts within $5\,r_{\mathrm{eff}}$ (i.e., for MRK\,1216 and PGC\,32873), we extracted CCD spectra and calculated model fluxes; otherwise, we used aperture photometry on the imaging data. {As the EPIC observations of the eFEDs targets are dominated by AGN emission and lack detectable diffuse emission (see Section \ref{sec:res-hr}, these sources are excluded from further analysis.}
 
At each galaxy's position, where the aperture contains fewer than $200$ net counts (NGC\,472 and PGC\,70520),  we performed photometric measurements of the $0.5-2$\,keV flux using circular apertures with a radius of $20 \, r_{\mathrm{eff}}$, while removing any overlapping point sources detected with \texttt{wavdetect}, i.e., those described in Section\,\ref{sec:res-ps}. Background was extracted from the same chip as the source emission. The flux-to-luminosity conversion was performed assuming APEC emission with $Z= 0.6\ \rm{Z_{\odot}}$ {\citep{1989GeCoA..53..197A}} and a temperature of $kT = 0.5\,\rm keV$.

In photometric measurements, the diffuse X-ray emission of galaxies is contaminated by the collective emission of unresolved low-mass X-ray binaries (LMXBs). Since LMXBs trace the stellar mass\,\citep{2004MNRAS.349..146G}, their contribution can be estimated for the present sample. Using the {LMXB luminosity function} for nearby galaxies \citep{2004MNRAS.349..146G}, and the stellar mass and distance of our targets, we estimated the luminosity contribution of unresolved LMXBs for both NGC\,472 and PGC\,70520, and applied the corresponding correction to the measured CGM luminosities (Table\,\ref{tab:measured}). Unresolved LMXBs are found to contribute, on average, roughly 40\% of the $0.5\text{--}2$~keV diffuse emission in these relatively faint X-ray galaxies.
Note that we did not consider the contribution from high-mass X-ray binaries, as it is expected to be insignificant in galaxies with old stellar populations.

For the two brightest sources (MRK\,1216 and PGC\,32873), we extracted CCD spectra from ACIS and EPIC data within a single aperture of $20\,r_{\rm eff}$, and calculated model fluxes.
We fit the galaxies' thermal emission spectrum with a \textit{phabs(apec+pow)} model {(in \texttt{Xspec} terminology)}, incorporating both the photon absorption, corresponding to the line-of-sight galactic hydrogen column density (Table\,\ref{tab:gal-prop}), and the contribution of the unresolved cosmic X-ray background with a photon index fixed at $\Gamma = 1.6$. 
The spectral fits were performed using \texttt{Xspec}. In the \textit{Chandra} ACIS fit, plasma temperature, abundance, and spectral normalizations were free parameters, whereas in the \textit{XMM-Newton} EPIC fit, the abundance was fixed at the best-fit ACIS value. {Note that allowing abundances to vary during the EPIC spectral fitting yields consistent results with those obtained using the fixed ACIS best-fit values.} For the ACIS spectrum of MRK\,1216 we applied a two-temperature model with the low-T component abundance linked to the abundance of the high-T component.
The fit range was set to $0.5-7$\,keV for ACIS and $0.5-10$\,keV for XMM-Newton spectra. A joint fit was applied for the MOS1 and MOS2 data, while the PN data was separately modeled. 
The source luminosities and the best-fit spectral model parameters are listed in Table\,\ref{tab:measured} and Table\,\ref{tab:spec-fit}, respectively.

The measured X-ray luminosities of diffuse emission, shown in Table\,\ref{tab:measured}, vary widely among galaxies, covering a luminosity range of $\gtrsim\!2$ orders of magnitude, highlighting the diversity of X-ray halos in the population of cE galaxies.
In particular, MRK\,1216 and PGC\,32873 display the highest luminosities in the sample with their X-ray halo extending beyond their effective radii (Section\,\ref{sec:sb-prof}, Figure\,\ref{fig:sbp}). On the other hand, NGC\,472 and PGC\,70520,  despite having comparable stellar masses and effective radii, lack prominent X-ray halos. We explore how these X-ray luminosities relate to local scaling relations in Section\,\ref{sec:discussion1}.

\begin{figure*}[hpt]
\centering
{\includegraphics[width=.245\textwidth]{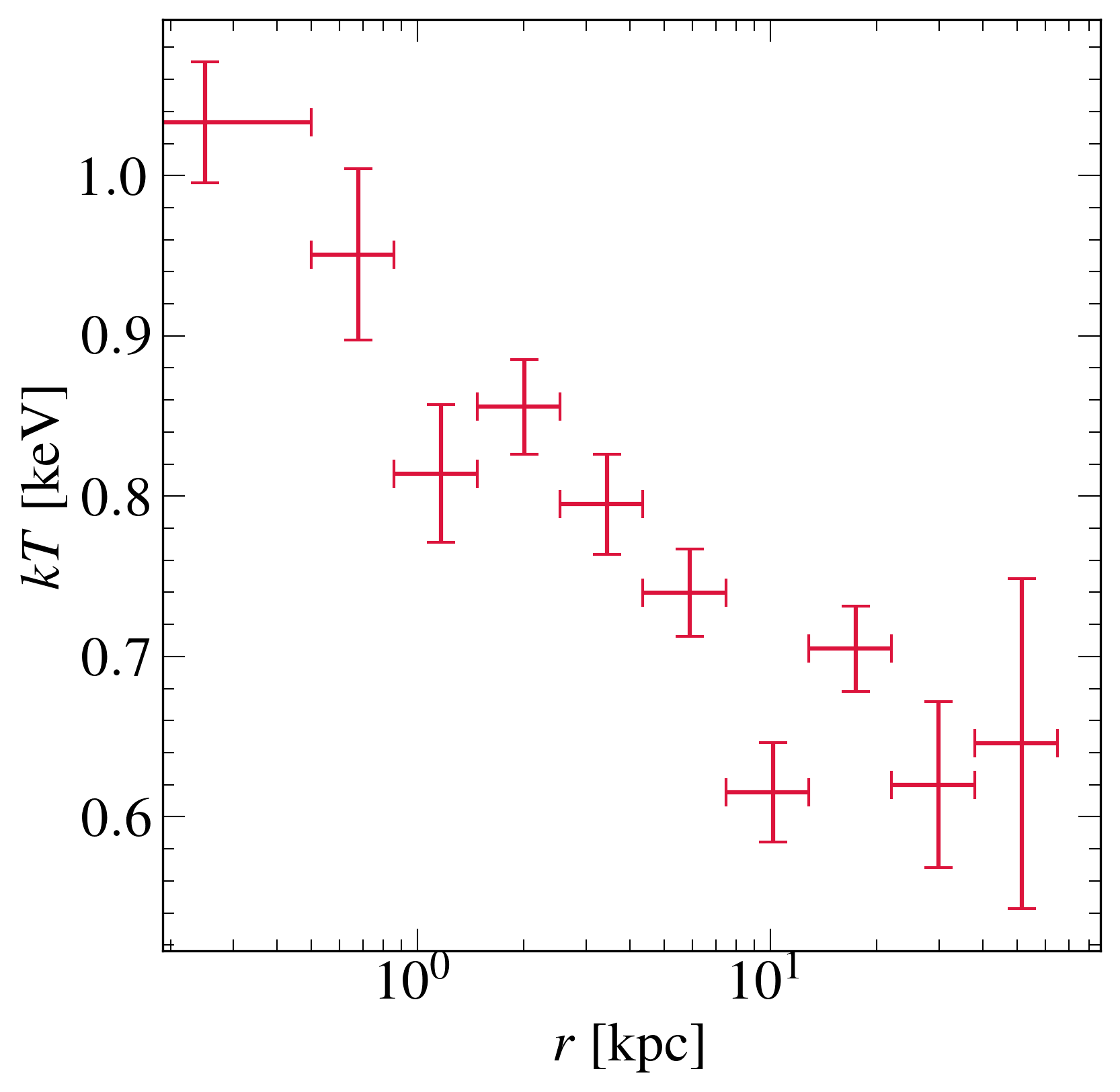}}
{\includegraphics[width=.249\textwidth]{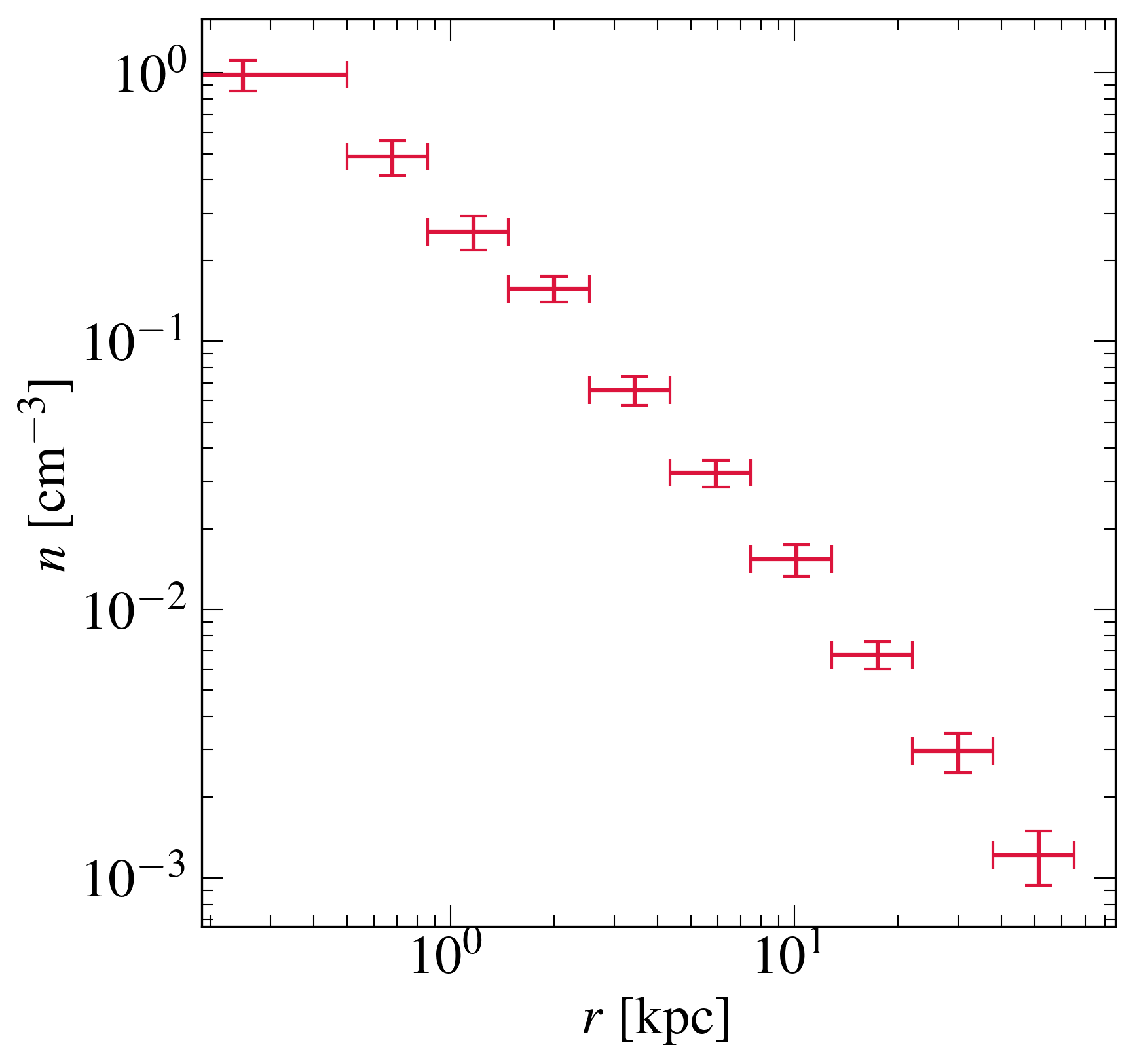}}
\vspace{0cm}
{\includegraphics[width=.243\textwidth]{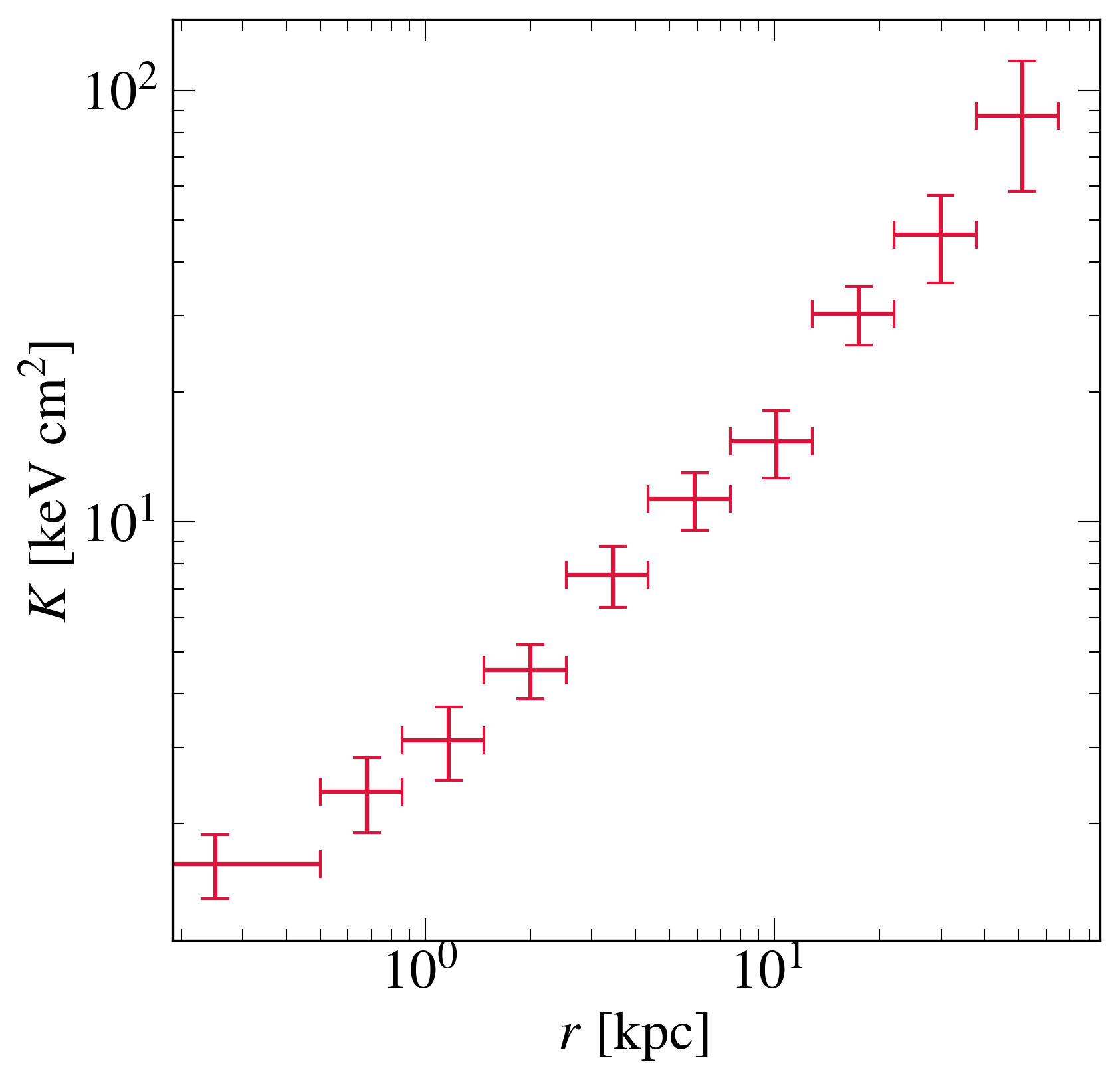}}
{\includegraphics[width=.249\textwidth]{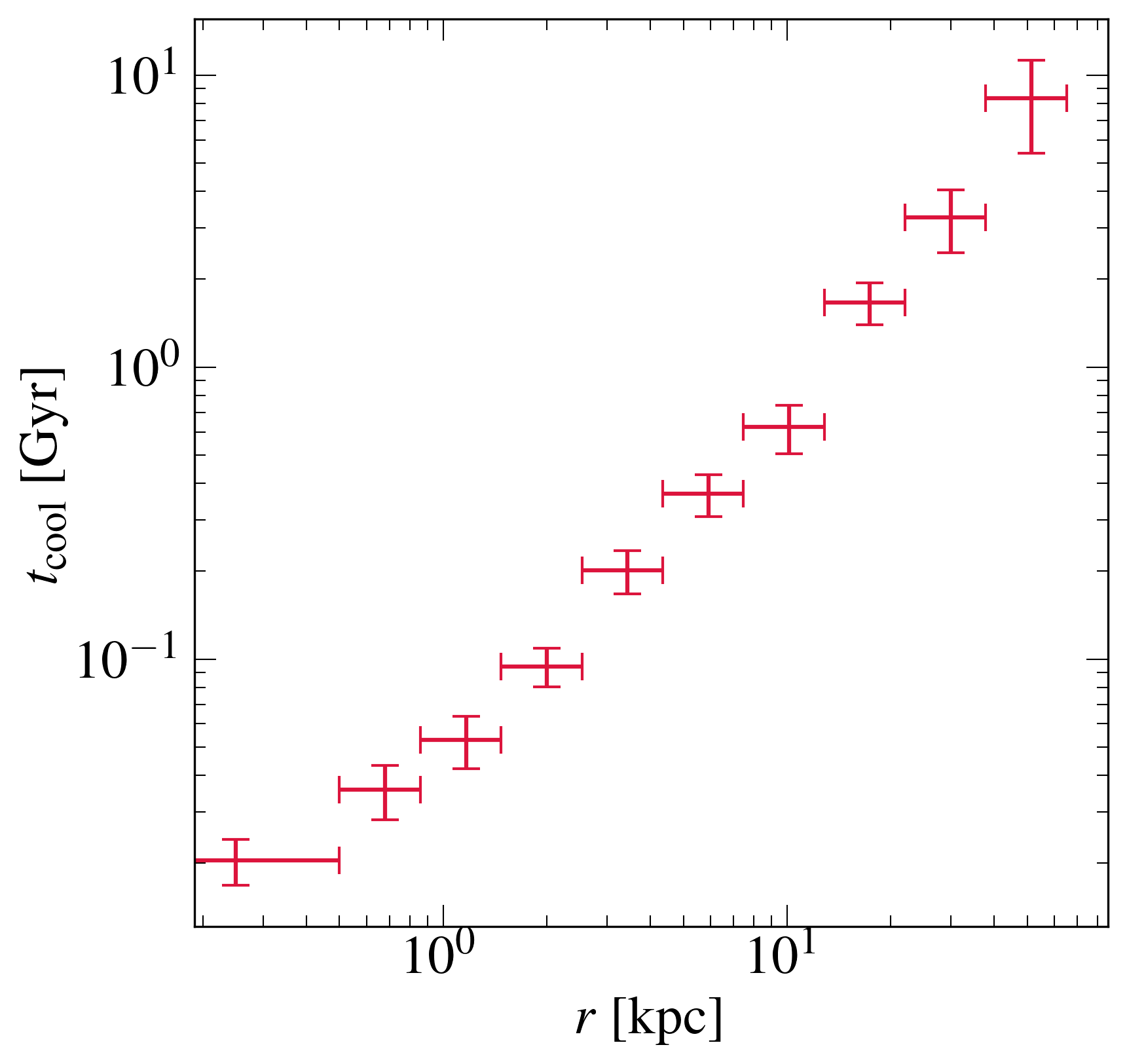}}
\vspace{0cm}
\caption{Deprojected thermodynamic profiles of the CGM surrounding MRK\,1216 extracted from \textit{Chandra} ACIS data, showing temperature ($kT$), total density ($n = n_e + n_i$), entropy ($K$), and cooling time ($t_{\mathrm{cool}}$) profiles extending out to $\gtrsim\!20 \, r_{\mathrm{eff}}$. {The effective radius of MRK\,1216 is $r_{\mathrm{eff}} = 3.0$\,kpc, corresponding to $6.8\arcsec$.}}
\label{fig:mrk1216-profiles}
\end{figure*}

\begin{figure*}[hpt]
\centering
{\includegraphics[width=.238\textwidth]{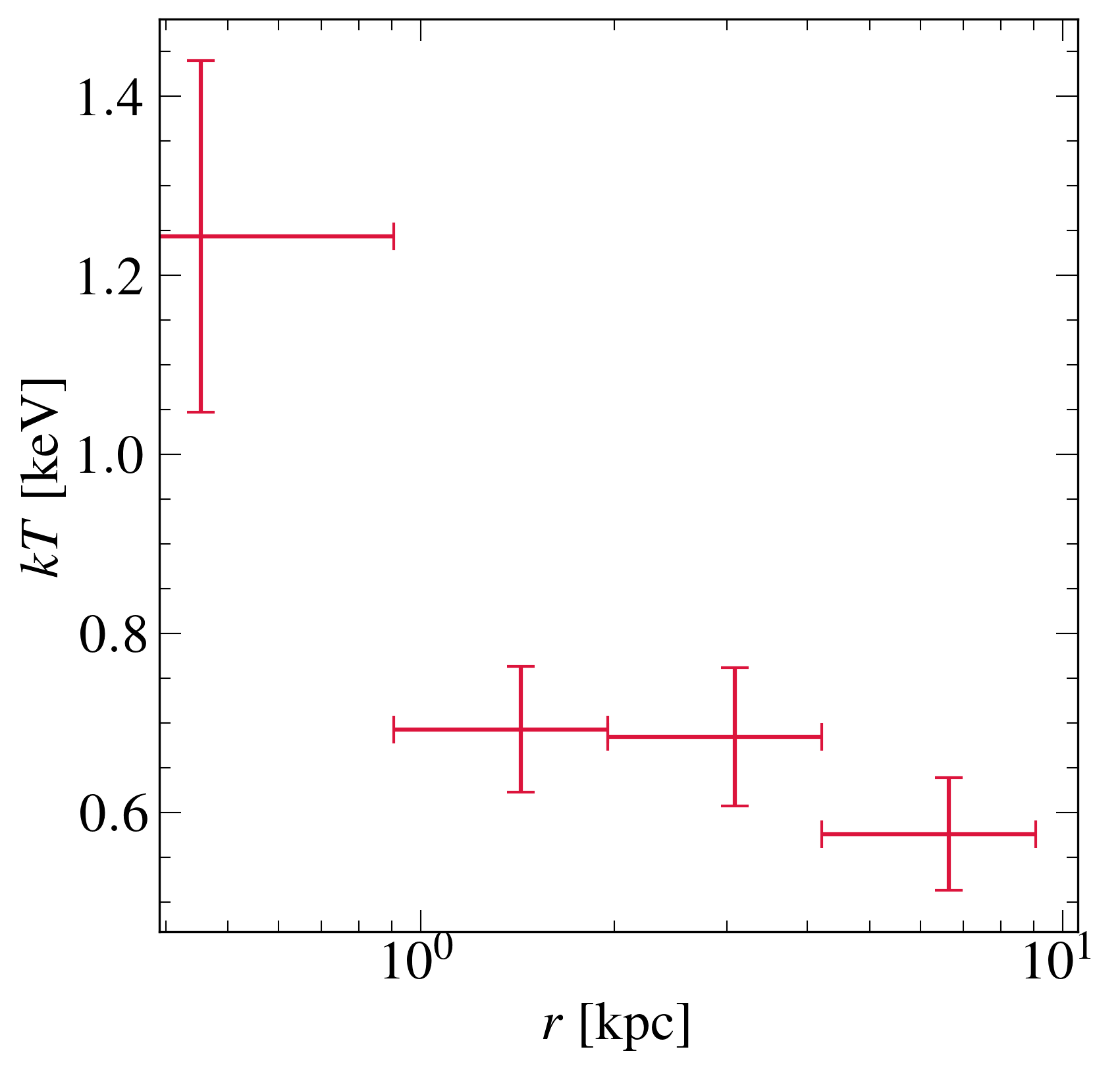}}
{\includegraphics[width=.246\textwidth]{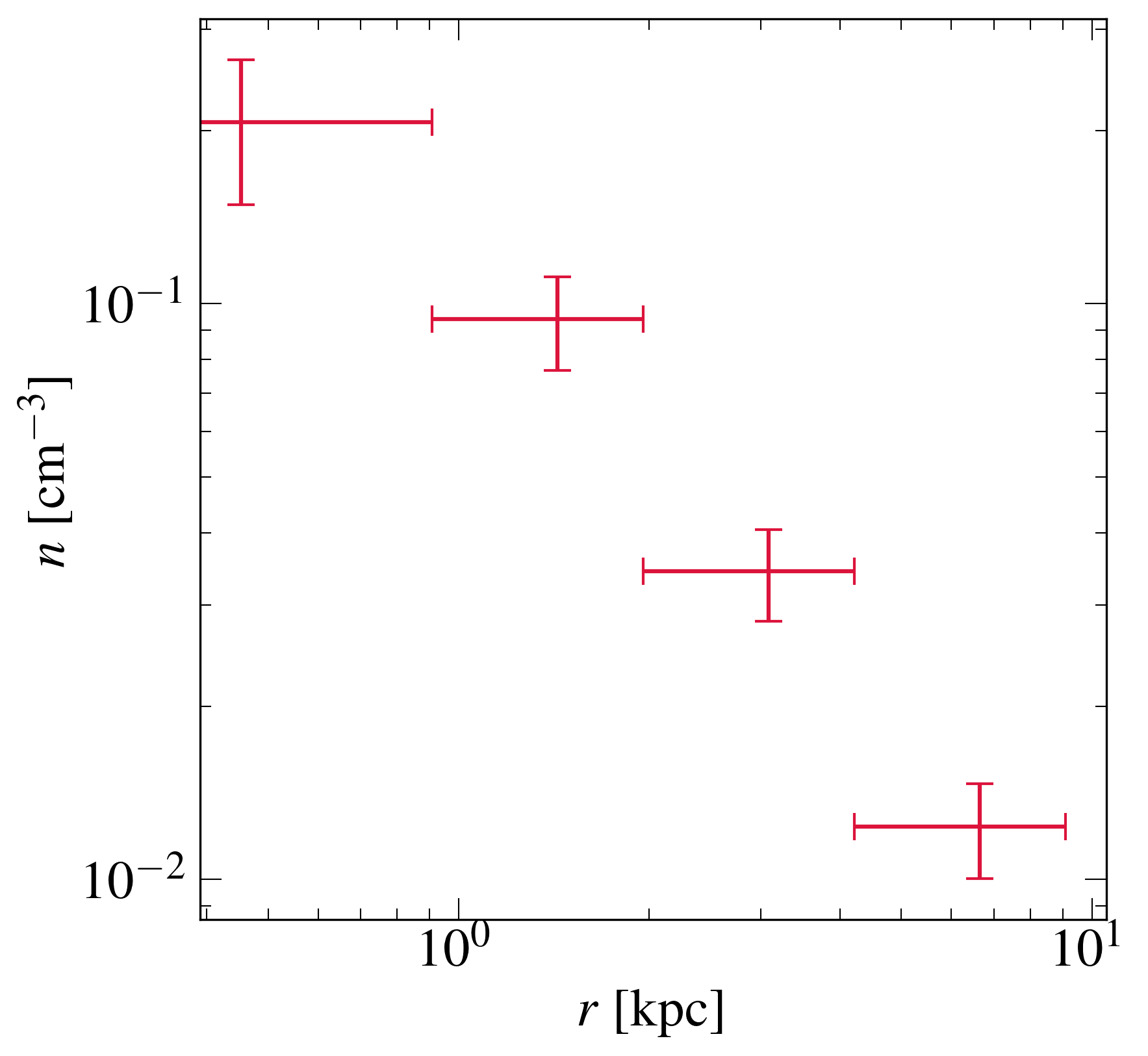}}
\vspace{0cm}
{\includegraphics[width=.258\textwidth]{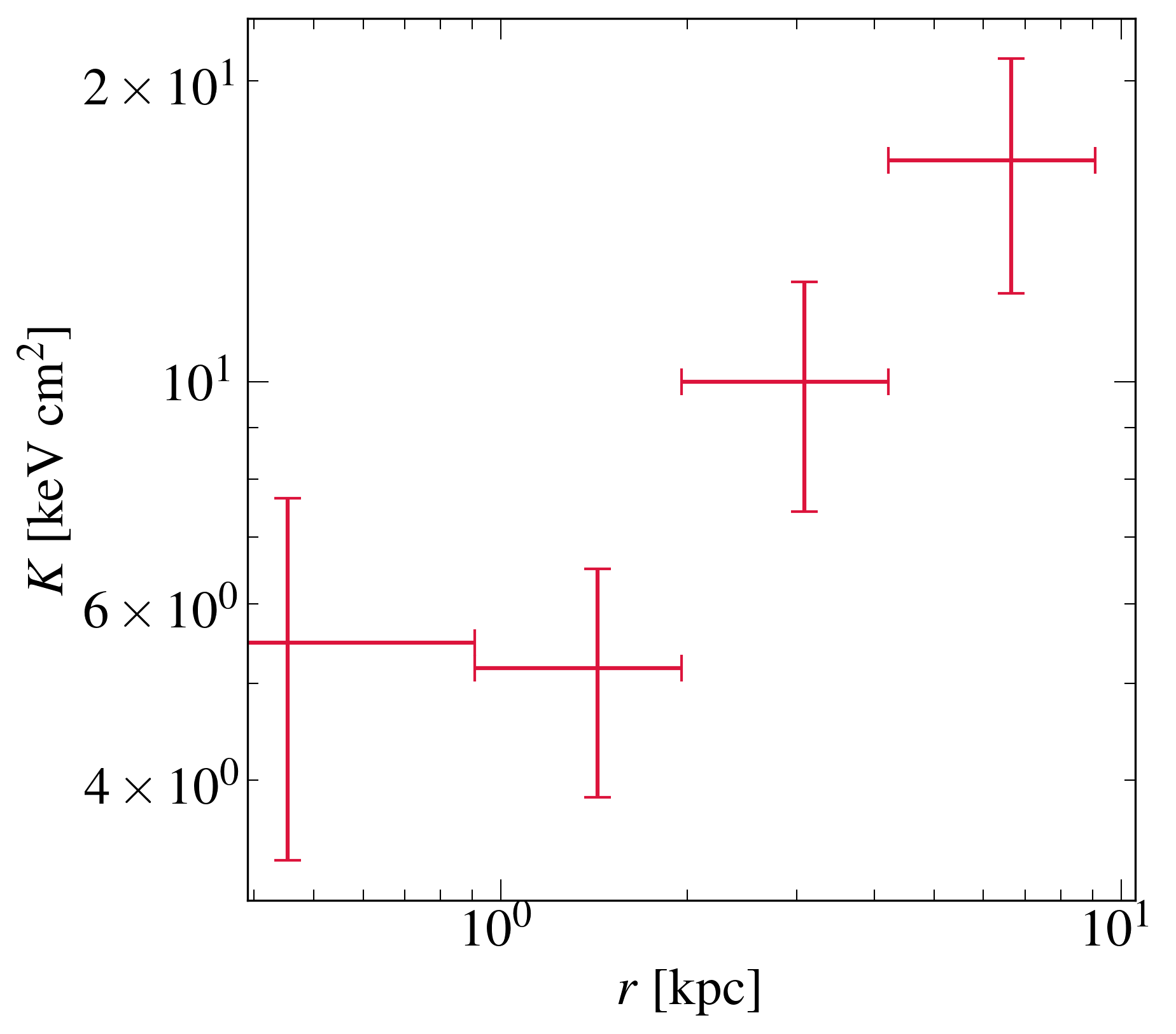}}
{\includegraphics[width=.244\textwidth]{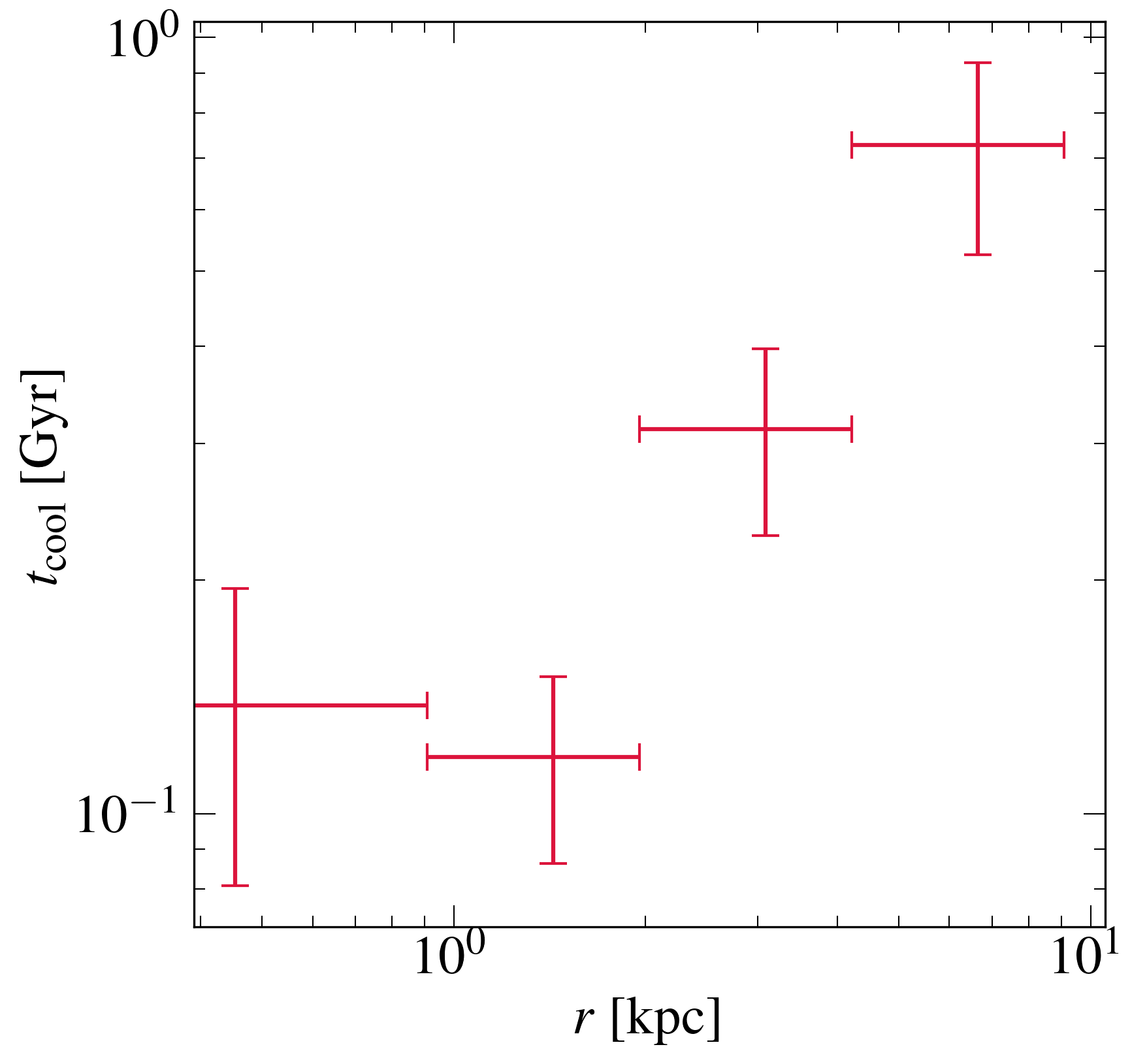}}
\vspace{0cm}
\caption{Deprojected thermodynamic profiles of the CGM surrounding PGC\,32873 extracted from \textit{Chandra} ACIS data, showing temperature ($kT$), total density ($n = n_e + n_i$), entropy ($K$), and cooling time ($t_{\mathrm{cool}}$) profiles extending out to $\sim\!3 \, r_{\mathrm{eff}}$. {The effective radius of PGC\,32873 is $r_{\mathrm{eff}} = 2.3$\,kpc, corresponding to $4.4\arcsec$.}}m
\label{fig:pgc32873-profiles}
\end{figure*}

\subsection{Surface brightness profile of MRK\,1216 and PGC\,32873} \label{sec:sb-prof}
We extracted the $0.3-2$\,keV radial surface brightness profiles for MRK\,1216 and PGC\,32873, the brightest sources of the sample.
The remaining galaxies had too few ($<50$) counts for this analysis.
Profile extraction and fitting were performed on ACIS data, with point sources removed, using the \texttt{pyproffit} software \citep{2020OJAp....3E..12E}. {The extracted surface brightness profiles were fitted with a single $\beta$-model:
$$I(r) = \mathrm{log}I_0 \left(1 + (x/r_c)^2\right) ^ {-3 \beta + 0.5} + \mathrm{log}B\,\,,$$
where $r_c$ is the core radius of the gas distribution and $B$ is the background level.}
The local background was estimated from an outer annulus centered on the galaxy. Given the Poissonian nature of the data, parameter estimation was performed using C-statistics. The background-subtracted profiles, along with the corresponding best-fit $\beta$-models, are displayed in Figure\,\ref{fig:sbp}. 

We trace the X-ray halos of MRK\,1216 and PGC\,32873 out to $\sim10,r_{\rm eff}$ and $\sim4,r_{\rm eff}$, respectively.
The best-fit core radius of the gas distribution is $0.874 \pm 0.001$\,kpc for MRK\,1216 and $0.492 \pm 0.001$\,kpc for PGC\,32873, with model slopes of $\beta = 0.507 \pm 0.003$ and $\beta = 0.526 \pm 0.011$, respectively. While these $\beta$ values are typical for bright ellipticals with $L_X \approx 10^{39}-10^{42} \, \rm erg\,s^{-1}$, the $r_{c}$ core radii are only a fraction of the typical $r_c \approx 2$\,kpc \citep[e.g.][]{1995A&A...298..784G}.

\subsection{Thermodynamic profiles} \label{sec:thermo-prof}
Following the method outlined in \cite{2023A&A...678A..91K} and \cite{2023A&A...678A.122Z}, we extracted X-ray spectra in concentric annuli from the ACIS data of MRK\,1216 and PGC\,32873, and derived the deprojected radial profiles of the thermodynamic properties of the CGM {assuming a model described in Section\,\ref{sec:res-de} with the abundances fixed to the value obtained from the fit within $10\,r_{\rm eff}$ (Table\,\ref{tab:spec-fit}).} The extraction annuli were defined on a logarithmic scale, with each radial bin containing $>\!200$ net counts. Due to its noncircular symmetry, elliptical annuli were used for PGC\,32873.
The thermodynamic profiles for MRK\,1216 and PGC\,32873 are displayed in Figure\,\ref{fig:mrk1216-profiles} and Figure\,\ref{fig:pgc32873-profiles}, respectively. Note that the outermost data points of these profiles derived from the deprojection may be subject to systematic uncertainties, potentially due to morphological asymmetries or residual CGM emission beyond the extraction radius.

We obtained the $kT$ temperature profiles directly from the best-fit $0.5-7$\,keV APEC model of each annulus, and derived the $n_e$ electron density from the model normalization, then expressed the $n = n_{\mathrm{e}} + n_{\mathrm{i}}$ total density assuming $n_{\mathrm{i}} = 0.92n_{\mathrm{e}}$ for the ion density. The $K = kT/n_{\mathrm{e}}^{2/3}$ entropy and the $t_{\mathrm{cool}} = 3/2(n_{\mathrm{e}} + n_{\mathrm{i}})kT / (n_{\mathrm{e}}  n_{\mathrm{i}} \Lambda(T))$ cooling time of the plasma were calculated from these deprojected properties, where $\Lambda(T)$ is the cooling function for solar metallicity{; this assumption may lead to a slight underestimate of the cooling time in regions with sub-solar abundance.}

The $kT$ profile of MRK\,1216 shows a smooth decline with radius with a core temperature of $1.03 \pm 0.04$\,keV, while PGC\,32873 displays a more centrally peaked distribution with a slightly higher core temperature of $1.24 \pm 0.20$\,keV. Similar central temperatures have been reported for giant ellipticals by \cite{2012MNRAS.425.2731W,2018MNRAS.481.4472L}. 
The central electron density of MRK\,1216, measured on a $0.5$\,kpc scale, is $0.51 \pm 0.13\, \mathrm{cm}^{-3}$. This value exceeds the $0.11 \pm 0.06\, \mathrm{cm}^{-3}$ central density measured for PGC\,32873 within $1$\,kpc, which is consistent with the typical $\sim\!0.1\,\mathrm{cm}^{-3}$ density observed in elliptical galaxies \citep{2003ARA&A..41..191M}. 
While both density profiles are smooth, the steeper gradient observed in MRK\,1216 points to a more concentrated hot gas distribution compared to PGC\,32873. The entropy profile of MRK\,1216 increases with the radius following $K \propto r^{0.88}$. This profile is broadly in line with, but slightly shallower than, the theoretical $K\propto r^{1.1}$ profile expected for systems where the hot atmosphere is shaped solely by gravitational forces \citep[e.g.][]{2005MNRAS.364..909V,2018ApJ...862...39B}. 
In contrast, PGC\,32873 shows a shallower entropy profile with a power-law slope of $\Gamma = 0.58$, and features a flattened core.
For MRK\,1216, we measure a cooling time of $t_{\mathrm{cool}} = 20.4 \pm 3.7 $\,Myr within $0.5\,\mathrm{kpc}$, and $t_{\mathrm{cool}} = 137.9 \pm 57.2$\,Myr for PGC\,32873 within $1\,\mathrm{kpc}$.

In \cite{2018MNRAS.477.3886W}, the authors provide a detailed explanation of the short central cooling time of MRK\,1216 linking the absence of star formation with radio-mechanical AGN feedback. While PGC\,32873's cooling time is slightly longer than that of MRK\,1216, it is still relatively short for a galaxy without ongoing star formation, indicating the presence of a heating mechanism in this system as well.
In addition, the flat central entropy distribution in PGC\,32873 and the high central temperature are also unmistakable signs of heating \citep[see e.g.][]{2012MNRAS.425.2731W,2018MNRAS.481.4472L}. In comparison, the central entropy distribution in MRK\,1216 is steeper and monotonously increasing, but its index is also flatter than $1.1$ expected for pure gravitational heating. In addition, the flattened central entropy profile of PGC\,32873 may reflect a more violent or more recent episode of AGN feedback compared to MRK\,1216. 
\textcolor{black}{Recently, \citet{fabian2023} tested a hidden (absorbed) cooling flow (HCF) model, and showed that if such a model applies, the implied cooling rate could be as high as $9.7\pm2.7$ $M_\odot$\,yr$^{-1}$. However, the current RGS data is not constraining enough to reliably determine the presence of a multi-temperature structure, such as the HCF model (see Section\,\ref{sec:hi-res-mrk}).}

\begin{table}[]
\caption{Best-fit model parameters for a single-temperature CIE model fit to the RGS spectra of MRK\,1216}
\label{tab:RGS}
\centering
\begin{tabular}{lc}
\hline\hline
$kT$ (keV) & $0.73\pm0.02$ \\  
$s$ & $0.53\pm0.04$ \\
O (Solar) & $1.3\pm0.4$ \\
Ne (Solar) & $1.4\pm0.5$ \\
Mg (Solar) & $2.1\pm0.7$ \\
Fe (Solar) & $1.0\pm0.3$ \\
\hline
\end{tabular}
\newline
\tablefoot{The spectra were extracted from a $0.5'$ wide region centred on MRK\,1216. The scale factor $s$ is the ratio of the observed LSF to the expected LSF based on the radial surface brightness profile extracted in the dispersion direction. The abundances are given with respect to the proto-Solar abundances of \citet{lodders2009}.}
\end{table}

\subsection{High-resolution spectrum of MRK\,1216} \label{sec:hi-res-mrk}
We modelled the RGS spectrum extracted from a 0.5 arcmin wide region with an absorbed collisionally ionised equilibrium (CIE) plasma model using the SPEX spectral fitting package version 3.08.01, which uses the atomic database SPEXACT 2.07.00 \citep{kaastra_2024_12771915} in the 8--21 \AA\ band. We first fitted a multi-temperature differential emission measure model with a Gaussian emission measure distribution \citep[see {\it gdem} in][]{deplaa2017}. However, the best-fit model did not provide a significant $\sigma$ parameter, and therefore, was consistent with a single temperature. 
The best-fit parameters of the single-temperature model are given in Table \ref{tab:RGS}. {We note that systematic uncertainties in abundances due to temperature structure are estimated to be of the order of $\sim\!20\%$ \citep{deplaa2017}, which remain smaller than the statistical uncertainties on our measured RGS abundances. Therefore, fitting with a multi-temperature model is not expected to significantly alter the abundance results.}

The RGS spectrum allows us to detect the abundances of O, Ne, Mg, and Fe at the $\sim\!3\sigma$ confidence level. The best-fit abundance ratios with respect to Fe hint at super-Solar values: $\mathrm{O/Fe} =1.3\pm0.6\, \rm Solar$, $\mathrm{Ne/Fe}=1.4\pm0.7\, \rm Solar$, and $\mathrm{Mg/Fe}=2.1\pm0.9\, \rm Solar$.

\begin{figure*}[hpt]
\centering
{\includegraphics[width=.45\textwidth]{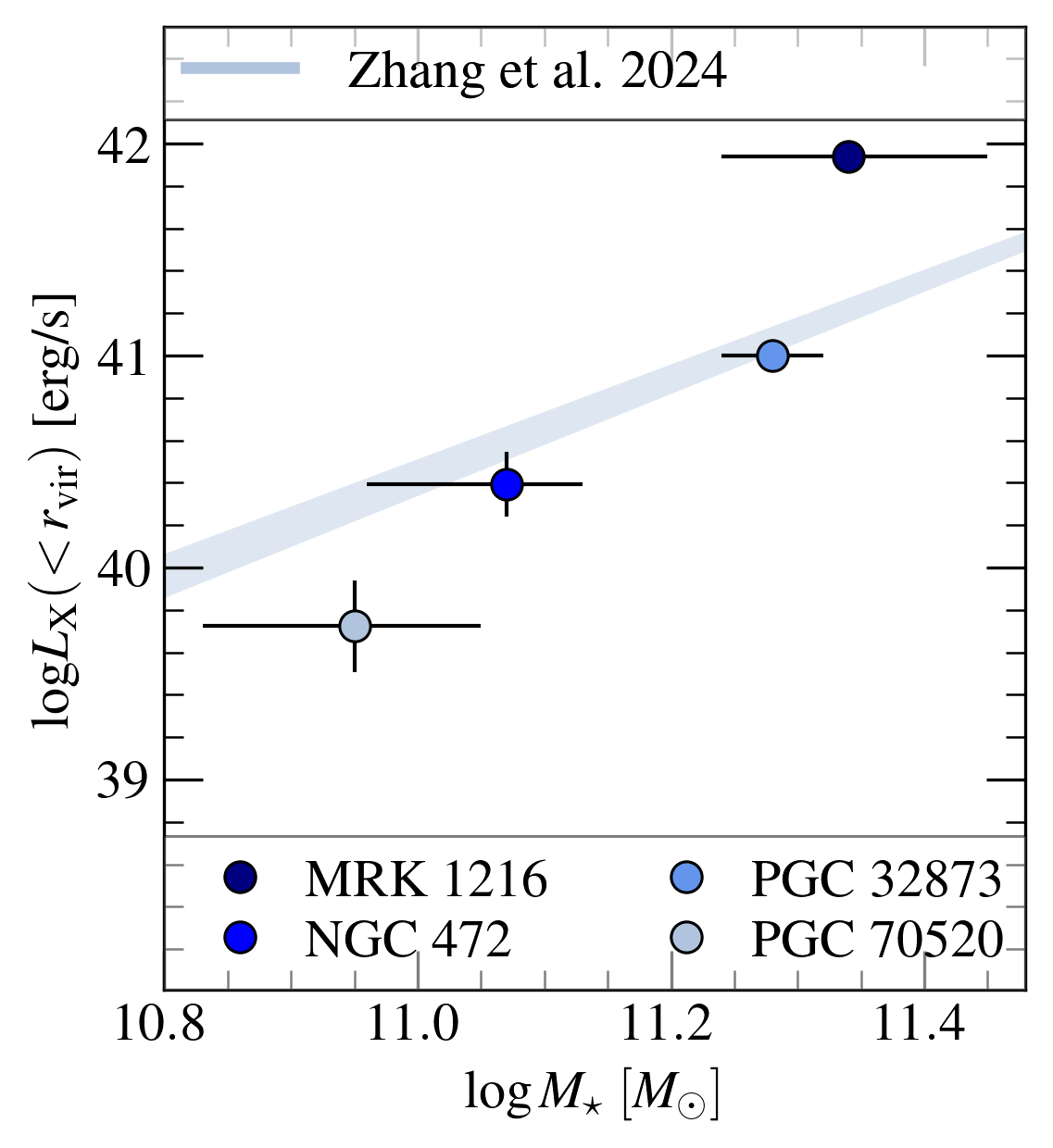}}
{\includegraphics[width=.48\textwidth]{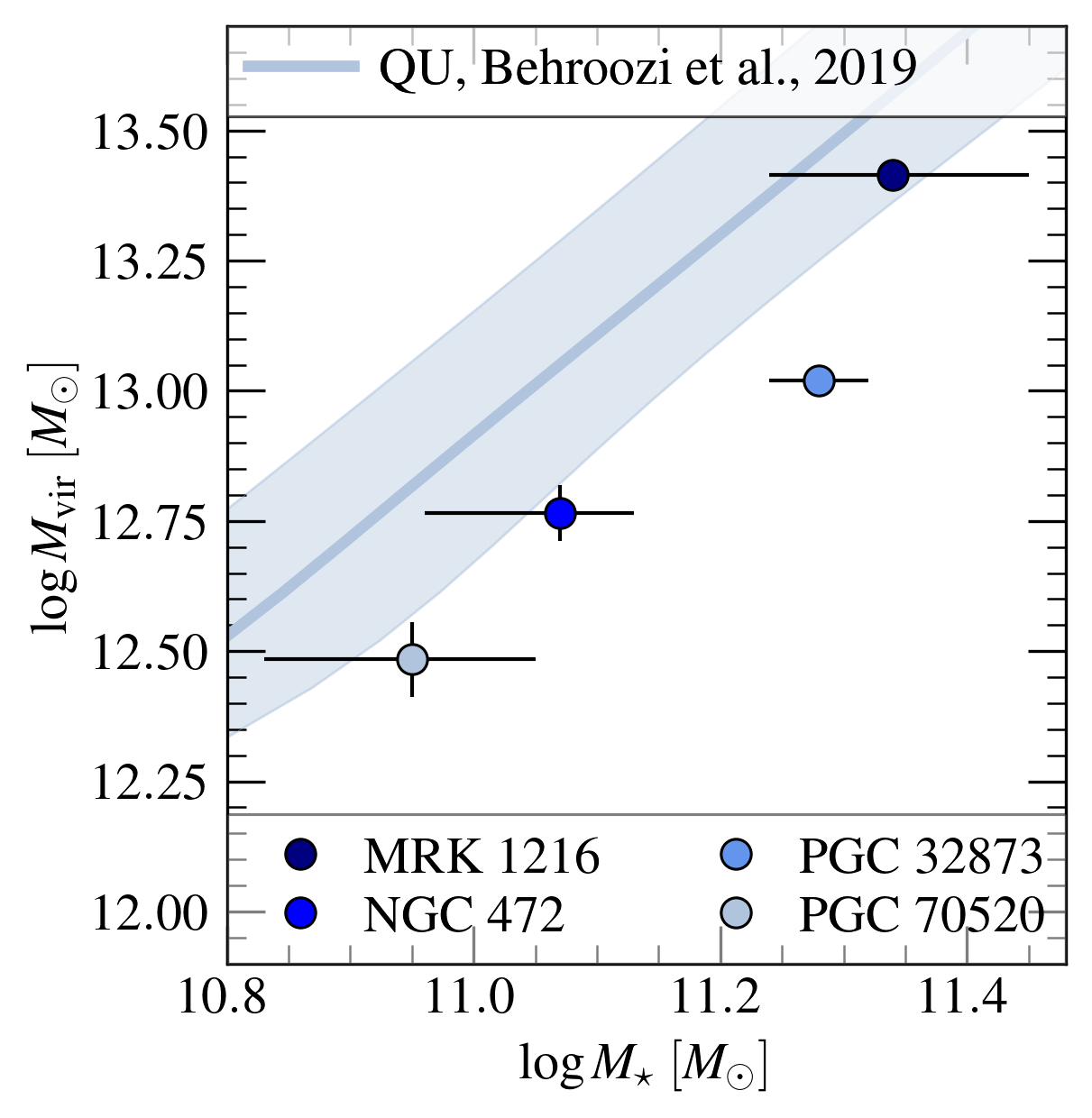}}
\caption{\textit{Left: }The $0.5-2$\,keV MOS1 CGM luminosity of the cE galaxies with detected X-ray emission within $r_{\mathrm{vir}}$ as a function of their stellar mass. The shaded region shows the best-fit relation and its $1\sigma$ uncertainty from \cite{2024A&A...690A.268Z}. \textit{Right: }The virial mass of cEs calculated from the $0.5-2$\,keV CGM luminosity as a function of stellar mass, and the best-fit relation at $z=0$ with its $1\sigma$ confidence interval from \cite{2019MNRAS.488.3143B}. Overall, based on X-ray measurements, the cEs largely follow both the $M_\star-L_X$ and the $M_\star-M_{\mathrm{vir}}$ relations established for the local galaxy population, however, their virial mass is on the lower side of expectations, possibly hinting at a more compact halo structure.} 
\label{fig:scalingrel}
\end{figure*}

\section{Discussion} \label{sec:discussion}
Out of the seven observed compact relic galaxies, MRK 1216 and PGC 32873 show significant extended X-ray emission originating from hot, diffuse atmospheres \citep[see also][]{2018MNRAS.477.3886W,2018ApJ...854..143B,2019ApJ...877...91B}. In comparison, NGC 472 and PGC 70520 are X-ray underluminous. The low X-ray luminosity might be the result of a lower halo mass, possibly combined with radio-mechanical AGN feedback blowing the hot gas out to larger radii. This diversity is reminiscent of the wide range of X-ray properties observed in the general population of local elliptical galaxies \citep[e.g.][]{2012ApJ...758...65B,2017ApJ...848...26G}. For the three galaxies only observed by XMM-Newton, the present data do not allow us to determine the presence or absence of hot X-ray emitting atmospheres.

The stellar population of MRK 1216 has, similarly to most giant ellipticals \citep{conroy2014}, high $\alpha$/Fe abundance ratios \citep{2017MNRAS.467.1929F}, and therefore, we would expect that the continuous, ongoing enrichment by thermalised stellar wind material from this isolated galaxy will also result in super-solar atmospheric $\alpha$/Fe abundance ratios. This would be contrary to the measurements of almost all giant ellipticals observed by {\it XMM-Newton} RGS \citep{deplaa2017,mernier2018,2019SSRv..215....5W} other than perhaps NGC4552 \citep{kara2024}. However, given that contrary to giant ellipticals, this galaxy evolved in relative isolation, it is plausible that the metal abundance ratios in the CGM of this galaxy will reflect the stellar abundances. Unfortunately, even though this observation provides an intriguing hint,  due to the large uncertainties, we are unable to make strong conclusions about the enrichment of this potentially ancient galactic atmosphere.

These X-ray observations of compact relic galaxies suggest that red nuggets observed at high redshifts \citep[e.g.][]{daddi2005} will also have diverse properties of X-ray atmospheres. Unfortunately, even deep observations with future missions with large effective areas, such as {\it NewAthena}, will not allow us to study the X-ray atmospheres of red nuggets at redshifts above $z\sim1$. Therefore, massive relic galaxies provide a rare and valuable opportunity to learn about the hot CGM in the early Universe.

\subsection{Comparison with the general galaxy population}
\label{sec:discussion1}
In this section, we compare key properties ($L_{\rm X}$, $M_{\rm vir}$, and $M_{\rm BH}$) of the cE galaxies with those of the broader galaxy population.

Figure\,\ref{fig:scalingrel} shows the $0.5-2$\,keV luminosity ($L_{\rm X}$) and the virial mass ($M_{\rm vir}$) as a function of $M_{\star}$ for the four galaxies in our sample with detected X-ray emission within $r_{\rm vir}$. 
To measure $L_{\mathrm X}$ within $r_{\mathrm{vir}}$, we first determined $r_{\mathrm{vir}}$ corresponding to $M_{\mathrm{vir}}$ from the $M_*-M_{\mathrm{vir}}$ relation of \cite{2019MNRAS.488.3143B}, then estimated the luminosity contribution between $20\,r_{\mathrm{eff}}$ and $r_{\mathrm{vir}}$, and added it to the luminosity measured within $20\,r_{\mathrm{eff}}$. 
The last step was necessary because the FOV and background level limit the maximum achievable source extraction radius.
The $20\,r_{\rm vir} - r_{\rm eff}$ luminosity contribution was calculated for a SB distribution described with a single $\beta$-model assuming $\beta=0.6$ \citep[e.g.][]{2003MNRAS.340.1375O,2018MNRAS.477.3886W}. Note that the actual $\beta$ parameter might vary depending on the specific properties of each galaxy (see Section\,\ref{sec:sb-prof}), but a value of $0.6$ is a reasonable assumption especially for galaxy outskirts, where the X-ray luminosity drops rapidly.
We found that the luminosity within the virial radius is dominated by the innermost regions as only $5-10\%$ of the total emission originates from the region between $20\,r_{\rm eff}$ and $r_{\rm vir}$. The $M_{\rm vir}$ was then derived from $L_{\rm X}$ using the correlation for elliptical galaxies from \cite{2006ApJ...652L..17M}. As shown in Figure\,\ref{fig:scalingrel}, the cE galaxies in our sample show a steeper $M_{\star}-L_{\rm X}$ distribution relative to the power-law trend observed for the CGM in the eROSITA All-Sky Survey \cite{2024A&A...690A.268Z}, and are slightly below the $M_{\star}-M_{\rm vir}$ relation established for quiescent galaxies \cite{2019MNRAS.488.3143B}. The latter could hint at a more concentrated dark matter halo structure in cE galaxies compared to the general galaxy population, although the small sample size prevents firm conclusions. It is also consistent with the relatively small $\beta$-model core radii we found for the surface brightness profile of MRK\,1216 and PGC\,32873 (Section\ref{sec:sb-prof}).
\textcolor{black}{Note that MRK\,1216 and PGC\,32873 have virial masses of $\gtrsim\,10^{13}\,M_{\odot}$, placing them at the low-mass end of the galaxy group mass distribution. Although located in relatively sparse regions with few and less massive companions (particularly MRK\,1216; see Section\,\ref{sec:sample}), they might be the dominant members of poor galaxy groups.}

\begin{figure}
\includegraphics[width=.46\textwidth]{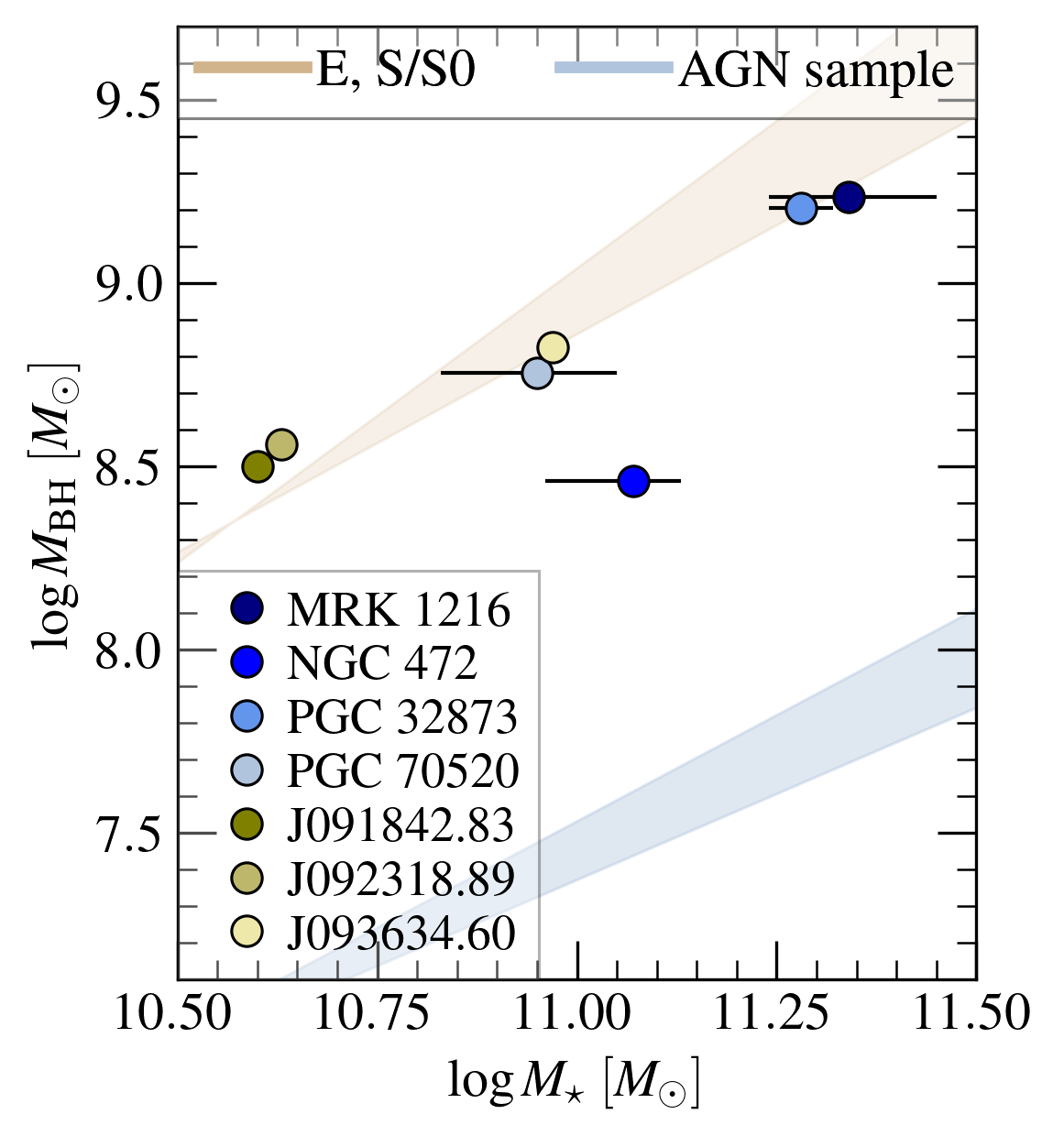}
\caption{The inferred black hole mass of cE galaxies, estimated from $\sigma$, as a function of stellar mass. The cEs in our sample follow the relation observed for inactive elliptical and spiral/S0 galaxies. For comparison, we also include the relation found for local AGNs. Both relations are taken from \cite{2015ApJ...813...82R}, with the shaded regions indicating the corresponding $1\sigma$ confidence intervals.}
\label{fig:mst-mbh}
\end{figure}

Figure\,\ref{fig:mst-mbh} shows the positions of the cE galaxies in the $M_{\star}-M_{\mathrm{BH}}$ plane, relative to the broader galaxy population.
$M_{\rm BH}$ is inferred from the $\sigma$ central velocity dispersion adopted from \cite{2017MNRAS.468.4216Y}. With the exception of NGC\,472, cE galaxies follow the relation established for galaxies hosting quiescent BHs, including elliptical galaxies and spiral/S0 galaxies with classical bulges. For reference, we also indicate the relation for AGN host galaxies \citep{2015ApJ...813...82R}. Interestingly, our cE galaxy sample includes four sources that may host AGN based on the present X-ray analysis, yet they still follow the trend of the inactive sample. Note, however, that the AGN sample used by \cite{2015ApJ...813...82R} to establish their relation was primarily identified through optical spectroscopy, specifically by detecting Seyfert-like narrow-line ratios and broad H$\alpha$ emission. While effective for classifying AGN in sources with strong emission lines, this method may miss heavily obscured or low-luminosity AGN, which could still be detectable in X-rays.

\subsection{Connection to little red dots}
\label{sec:discussion2}

\begin{figure}
\includegraphics[width=.49\textwidth]{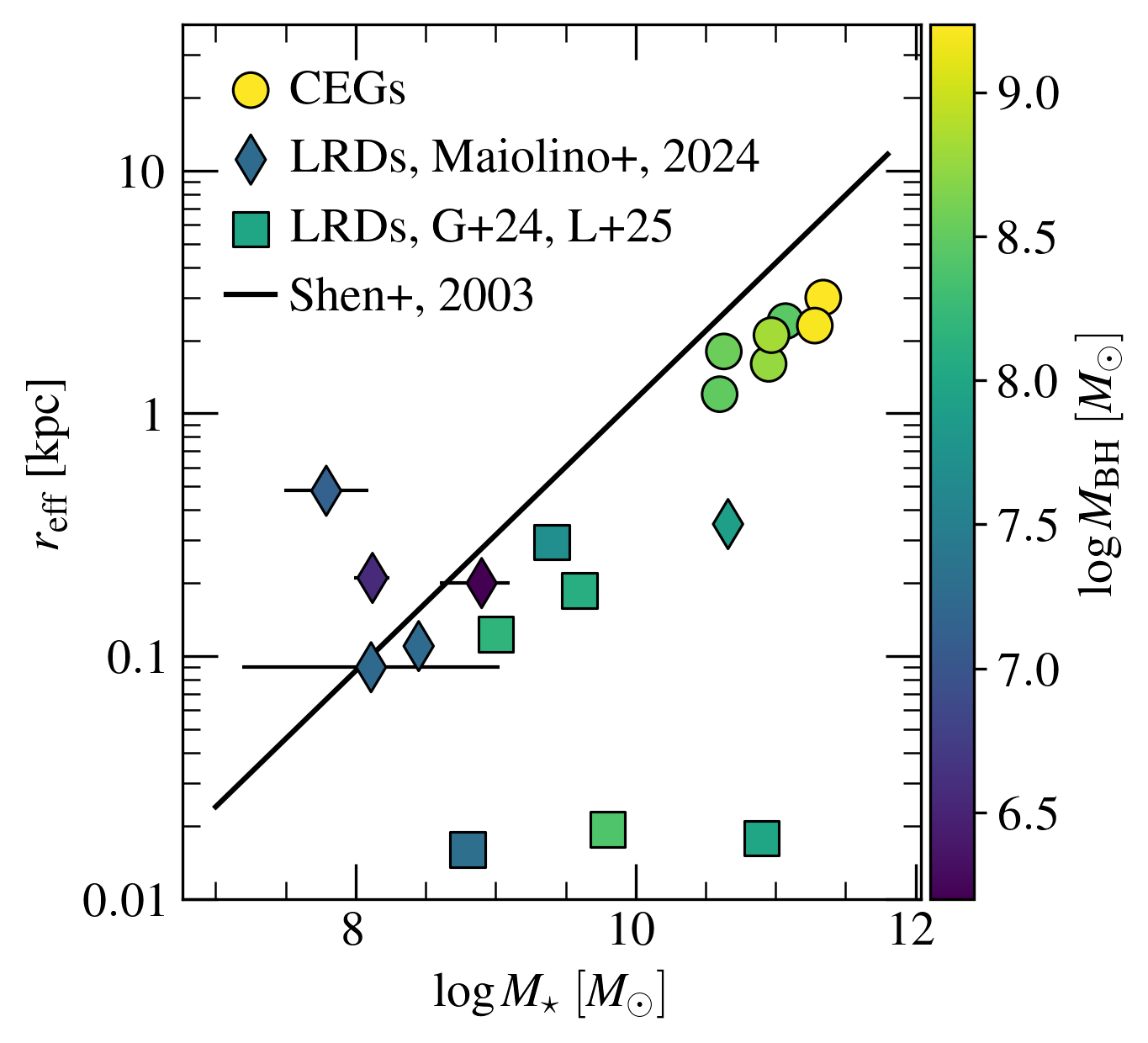}
\caption{Effective radius as a function of stellar mass for different galaxy samples (circles: cEs, diamonds: LRDs from \cite{2024A&A...691A.145M}, squares: LRDs with their size and stellar mass adopted from \cite{2025ApJ...978...92L}, and their black hole mass adopted from \cite{2024ApJ...964...39G}). The color of each data point represents the black hole mass as indicated by the color bar. The solid black line shows the relation established by \cite{2003MNRAS.343..978S} for local galaxies.}
\label{fig:mst-reff}
\end{figure}

Little red dots (LRDs) are a recently discovered class of galaxies identified in \textit{JWST} observations at high redshift, predominantly residing at $z = 3-7$. These galaxies are extremely compact, with effective radii on the order of $\sim\!100$ pc \citep{2023ApJ...955L..12B}, and exhibit a distinctive V-shaped spectral energy distribution (SED). Their unusual properties could be explained by either an exceptionally dense stellar population or the presence of a supermassive black hole (SMBH) that is potentially over-massive relative to the stellar mass of its host galaxy when compared to the local $M_{\rm BH} - M_{\rm \star}$ scaling relation. LRDs have stellar masses of $\sim 10^9 M_{\odot}$ and black hole masses ranging from $10^6$ to $10^8 M_{\odot}$ \citep[e.g.][]{matthee2024,2025ApJ...986..126K}.  

Despite numerous studies investigating their nature, the origin of LRDs remains debated. Observations report broad H$\alpha$ emission lines, suggesting the presence of an AGN, yet these galaxies are X-ray weak and lack significant dust, which is typically associated with star formation \citep{2024ApJ...969L..18A}. Additionally, LRDs appear to be relatively common in the early universe, with a number density of $10^{-4.0\pm0.1}$ Mpc$^{-3}$, accounting for $14\%\pm3\%$ of galaxies with similar masses at $z\approx6.9$ \citep{2024ApJ...968....4P}. Given their unique properties and high abundance, LRDs may represent the cores of $z\sim2$ red nuggets and the progenitors of today's massive elliptical galaxies, as we discuss below.  

Red nugget (RN) galaxies, typically observed at $z\!\sim\!2$, are more extended than LRDs, with effective radii of $\sim\!1$ kpc, and more massive, with $M_{\rm \star} \sim 10^{11} \ \rm{M_{\odot}}$. However, they remain significantly more compact than local elliptical galaxies and---like LRDs---lie below the $M_{\rm \star} - r_{\rm eff}$ relation (Figure \ref{fig:mst-reff}). Based on this observational evidence, we propose that RNs may have evolved from even more compact progenitors, namely LRDs.  

To assess the viability of this scenario, we estimate the number density of LRDs that could evolve into RNs based on the following considerations. For an LRD to be a viable progenitor of an RN, it must undergo at most a couple of major mergers, as such events could strip LRDs of their distinctive characteristics, particularly their compactness, which resembles that of RN galaxies. Recent studies also suggest that the decline in LRD occurrence from $z\approx5$ is due to mergers erasing their distinctive features \citep{2025ApJ...988L..22I,2025ApJ...986..126K}. In addition, based on N-body simulations, the core-scouring effect of supermassive black hole binaries puffs up an initially dense central core \citep{2009ApJ...699L.178N}. While minor mergers contribute to building up a galaxy's outer regions while preserving its dense core, multiple minor mergers ultimately lead to a more extended galaxy than a few major mergers of the same total stellar mass, making them less efficient at driving mass growth \citep{2009ApJ...699L.178N,2009ApJ...697.1290B,2024MNRAS.535.1202R}. Notably, LRDs must grow by approximately two orders of magnitude in both effective radius and stellar mass to acquire characteristics similar to those of RNs. Like the $z \sim 2$ red nuggets, Mrk\,1216 and PGC\,032873 exhibit strong rotation and highly peaked velocity dispersion profiles \citep{2017MNRAS.467.1929F}. However, multiple major mergers can lead to a loss of rotational support \citep{2009ApJ...699L.178N}.  

These considerations strongly suggest that mergers alone are unlikely to be the primary mechanism for transforming LRDs into RNs. Thus, we estimate the probability of LRDs preserving their distinctive features. We assume an initial LRD number density of $10^{-4} \,\rm Mpc^{-3}$ at $z \approx 6.9$, adopted from \cite{2024ApJ...968....4P}. Our estimate spans the period from $z = 6.9$ to $z = 1$, where the number density of RNs is well established. Over this redshift range (covering 5.36 Gyr), a typical galaxy experiences $\sim0.7$ major mergers per Gyr \citep{2024arXiv241104944D}.
Assuming a Poissonian process, the probability of an LRD experiencing one major merger for 5.36 Gyr is $\sim10\%$, yielding a surviving number density of  $\sim1.0 \times 10^{-5}\,\rm Mpc^{-3}$. Comparing this to the RN number density at the same redshift ($9.8 \times 10^{-6}\,\rm Mpc^{-3}$, \cite{2023A&A...669A..95L}), we find that LRDs could fully account for the observed red nugget population.  

In this context, the hidden (by cold absorption) cooling flow rate of $\sim10\, M_\odot$ yr$^{-1}$, which is larger than the rate inferred in typical early-type galaxies \citep{fabian2023}, together with the observed bottom-heavy initial mass function \citep[IMF,][]{2015ApJ...808...79F}, may provide another important clue about galaxy formation. It may indicate that the high thermal pressure in early cooling flows could have led to a bottom-heavy IMF, resulting in the accumulation of low-mass stars in the progenitors of present day massive galaxies \citep{fabian2024}.

In connecting cEGs with the LRD population, it is also worth noting the so-called green pea galaxies — compact, star-forming systems at lower redshifts — which may represent local analogs to certain phases in the evolution of LRDs \citep{2019ApJ...880..144S,lin2024discoverylocalanalogsjwsts}.

While current simulations struggle to reproduce the LRD population due to limitations in black hole seeding and stellar feedback modeling, future simulations incorporating more sophisticated black hole growth and star formation physics may offer a better understanding of their origins.

\section{Summary} \label{sec:summary}
This study presents the X-ray analysis of seven local compact elliptical galaxies (cEGs), selected for their morphological similarity to high-redshift red nuggets.
Using \textit{Chandra} and \textit{XMM-Newton} data, our analysis focuses on the properties of the hot gaseous halos surrounding these galaxies. The key findings of our study are as follows.
\begin{itemize}
\item The observed systems exhibit diverse X-ray halo properties, with $0.5$–$2$\,keV luminosities spanning $\gtrsim\!2$ orders of magnitude among MRK\,1216, NGC\,472, PGC\,32873, and PGC\,70520; for the remaining galaxies, observed only with \textit{XMM-Newton} EPIC, diffuse emission could not be resolved due to a possible central AGN. These results suggest that red nuggets at high redshift might also have diverse X-ray halo properties.
\item MRK\,1216 and PGC\,32873, the most luminous sources in the sample, exhibit extended X-ray halos reaching $\sim\!10\,r_{\rm{eff}}$ and $\sim\!4\,r_{\rm{eff}}$, respectively, with corresponding $0.5-2$\,keV luminosities of $\sim\!9\times10^{41}\, \rm erg \, s^{-1}$ and $\sim\!10^{41}\, \rm erg \, s^{-1}$.
\item NGC\,472 and PGC\,70520 appear X-ray underluminous. Their low X-ray luminosities may reflect lower halo masses, potentially coupled with radio-mode AGN feedback expelling hot gas to larger radii.
\item Deprojected thermodynamic profiles for MRK\,1216 and PGC\,32873 suggest AGN activity, as indicated by central temperature peaks, short central cooling times, and flat entropy slopes.
\item High-resolution RGS spectroscopy of MRK\,1216 hints at super-solar O/Fe, Ne/Fe, and Mg/Fe abundance ratios.
\item The cE galaxies exhibit a steeper $M_{\star}$–$L_{\rm X}$ relation compared to the CGM trend from the eROSITA All-Sky Survey, and lie slightly below the $M_{\star}$–$M_{\rm vir}$ relation for quiescent galaxies, indicating more concentrated dark matter halos.
\item The cE galaxies, with the exception of NGC\,472, follow the $M_{\star}$–$M_{\rm BH}$ relation established for quiescent galaxies.
\item cEGs lie below the $M_{\star}$–$r_{\rm eff}$ relation, similar to little red dot galaxies, hinting at a possible evolutionary connection.
\end{itemize}

\begin{acknowledgements}
O. E. K. and N. W. were supported by the GACR EXPRO grant No. 21-13491X. \'A. B. acknowledges support from the Smithsonian Institution and the Chandra Project through NASA contract NAS8-03060. The research leading to these results has received funding from the European Union’s Horizon 2020 Programme under the AHEAD2020 project (grant agreement n. 871158).
\end{acknowledgements}

\bibliographystyle{aa}
\bibliography{compact_elliptical_galaxies}

\begin{thebibliography}{74}
\expandafter\ifx\csname natexlab\endcsname\relax\def\natexlab#1{#1}\fi

\bibitem[{{Ananna} {et~al.}(2024){Ananna}, {Bogd{\'a}n}, {Kov{\'a}cs}, {Natarajan}, \& {Hickox}}]{2024ApJ...969L..18A}
{Ananna}, T.~T., {Bogd{\'a}n}, {\'A}., {Kov{\'a}cs}, O.~E., {Natarajan}, P., \& {Hickox}, R.~C. 2024, \apjl, 969, L18

\bibitem[{{Anders} \& {Grevesse}(1989)}]{1989GeCoA..53..197A}
{Anders}, E. \& {Grevesse}, N. 1989, \gca, 53, 197

\bibitem[{{Arnaud}(1996)}]{1996ASPC..101...17A}
{Arnaud}, K.~A. 1996, in Astronomical Society of the Pacific Conference Series, Vol. 101, Astronomical Data Analysis Software and Systems V, ed. G.~H. {Jacoby} \& J.~{Barnes}, 17

\bibitem[{{Babyk} {et~al.}(2018){Babyk}, {McNamara}, {Nulsen}, {Russell}, {Vantyghem}, {Hogan}, \& {Pulido}}]{2018ApJ...862...39B}
{Babyk}, I.~V., {McNamara}, B.~R., {Nulsen}, P.~E.~J., {et~al.} 2018, \apj, 862, 39

\bibitem[{{Baggen} {et~al.}(2023){Baggen}, {van Dokkum}, {Labb{\'e}}, {Brammer}, {Miller}, {Bezanson}, {Leja}, {Wang}, {Whitaker}, {Suess}, \& {Nelson}}]{2023ApJ...955L..12B}
{Baggen}, J. F.~W., {van Dokkum}, P., {Labb{\'e}}, I., {et~al.} 2023, \apjl, 955, L12

\bibitem[{{Behroozi} {et~al.}(2019){Behroozi}, {Wechsler}, {Hearin}, \& {Conroy}}]{2019MNRAS.488.3143B}
{Behroozi}, P., {Wechsler}, R.~H., {Hearin}, A.~P., \& {Conroy}, C. 2019, \mnras, 488, 3143

\bibitem[{{Bell} {et~al.}(2003){Bell}, {McIntosh}, {Katz}, \& {Weinberg}}]{2003ApJS..149..289B}
{Bell}, E.~F., {McIntosh}, D.~H., {Katz}, N., \& {Weinberg}, M.~D. 2003, \apjs, 149, 289

\bibitem[{{Bezanson} {et~al.}(2009){Bezanson}, {van Dokkum}, {Tal}, {Marchesini}, {Kriek}, {Franx}, \& {Coppi}}]{2009ApJ...697.1290B}
{Bezanson}, R., {van Dokkum}, P.~G., {Tal}, T., {et~al.} 2009, \apj, 697, 1290

\bibitem[{{Bogd{\'a}n} {et~al.}(2017){Bogd{\'a}n}, {Bourdin}, {Forman}, {Kraft}, {Vogelsberger}, {Hernquist}, \& {Springel}}]{2017ApJ...850...98B}
{Bogd{\'a}n}, {\'A}., {Bourdin}, H., {Forman}, W.~R., {et~al.} 2017, \apj, 850, 98

\bibitem[{{Bogd{\'a}n} {et~al.}(2012{\natexlab{a}}){Bogd{\'a}n}, {David}, {Jones}, {Forman}, \& {Kraft}}]{2012ApJ...758...65B}
{Bogd{\'a}n}, {\'A}., {David}, L.~P., {Jones}, C., {Forman}, W.~R., \& {Kraft}, R.~P. 2012{\natexlab{a}}, \apj, 758, 65

\bibitem[{{Bogd{\'a}n} {et~al.}(2012{\natexlab{b}}){Bogd{\'a}n}, {Forman}, {Zhuravleva}, {Mihos}, {Kraft}, {Harding}, {Guo}, {Li}, {Churazov}, {Vikhlinin}, {Nulsen}, {Schindler}, \& {Jones}}]{2012ApJ...753..140B}
{Bogd{\'a}n}, {\'A}., {Forman}, W.~R., {Zhuravleva}, I., {et~al.} 2012{\natexlab{b}}, \apj, 753, 140

\bibitem[{{Bogd{\'a}n} {et~al.}(2024){Bogd{\'a}n}, {Goulding}, {Natarajan}, {Kov{\'a}cs}, {Tremblay}, {Chadayammuri}, {Volonteri}, {Kraft}, {Forman}, {Jones}, {Churazov}, \& {Zhuravleva}}]{2024NatAs...8..126B}
{Bogd{\'a}n}, {\'A}., {Goulding}, A.~D., {Natarajan}, P., {et~al.} 2024, Nature Astronomy, 8, 126

\bibitem[{{Buitrago} {et~al.}(2018){Buitrago}, {Ferreras}, {Kelvin}, {Baldry}, {Davies}, {Angthopo}, {Khochfar}, {Hopkins}, {Driver}, {Brough}, {Sabater}, {Conselice}, {Liske}, {Holwerda}, {Bremer}, {Phillipps}, {L{\'o}pez-S{\'a}nchez}, \& {Graham}}]{2018A&A...619A.137B}
{Buitrago}, F., {Ferreras}, I., {Kelvin}, L.~S., {et~al.} 2018, \aap, 619, A137

\bibitem[{{Buote} \& {Barth}(2018)}]{2018ApJ...854..143B}
{Buote}, D.~A. \& {Barth}, A.~J. 2018, \apj, 854, 143

\bibitem[{{Buote} \& {Barth}(2019)}]{2019ApJ...877...91B}
{Buote}, D.~A. \& {Barth}, A.~J. 2019, \apj, 877, 91

\bibitem[{{Conroy} {et~al.}(2014){Conroy}, {Graves}, \& {van Dokkum}}]{conroy2014}
{Conroy}, C., {Graves}, G.~J., \& {van Dokkum}, P.~G. 2014, \apj, 780, 33

\bibitem[{{Daddi} {et~al.}(2005){Daddi}, {Renzini}, {Pirzkal}, {Cimatti}, {Malhotra}, {Stiavelli}, {Xu}, {Pasquali}, {Rhoads}, {Brusa}, {di Serego Alighieri}, {Ferguson}, {Koekemoer}, {Moustakas}, {Panagia}, \& {Windhorst}}]{daddi2005}
{Daddi}, E., {Renzini}, A., {Pirzkal}, N., {et~al.} 2005, \apj, 626, 680

\bibitem[{{Damjanov} {et~al.}(2009){Damjanov}, {McCarthy}, {Abraham}, {Glazebrook}, {Yan}, {Mentuch}, {Le Borgne}, {Savaglio}, {Crampton}, {Murowinski}, {Juneau}, {Carlberg}, {J{\o}rgensen}, {Roth}, {Chen}, \& {Marzke}}]{2009ApJ...695..101D}
{Damjanov}, I., {McCarthy}, P.~J., {Abraham}, R.~G., {et~al.} 2009, \apj, 695, 101

\bibitem[{{Davis} {et~al.}(2012){Davis}, {Bautz}, {Dewey}, {Heilmann}, {Houck}, {Huenemoerder}, {Marshall}, {Nowak}, {Schattenburg}, {Schulz}, \& {Smith}}]{2012SPIE.8443E..1AD}
{Davis}, J.~E., {Bautz}, M.~W., {Dewey}, D., {et~al.} 2012, in Society of Photo-Optical Instrumentation Engineers (SPIE) Conference Series, Vol. 8443, Space Telescopes and Instrumentation 2012: Ultraviolet to Gamma Ray, ed. T.~{Takahashi}, S.~S. {Murray}, \& J.-W.~A. {den Herder}, 84431A

\bibitem[{{de Plaa} {et~al.}(2017){de Plaa}, {Kaastra}, {Werner}, {Pinto}, {Kosec}, {Zhang}, {Mernier}, {Lovisari}, {Akamatsu}, {Schellenberger}, {Hofmann}, {Reiprich}, {Finoguenov}, {Ahoranta}, {Sanders}, {Fabian}, {Pols}, {Simionescu}, {Vink}, \& {B{\"o}hringer}}]{deplaa2017}
{de Plaa}, J., {Kaastra}, J.~S., {Werner}, N., {et~al.} 2017, \aap, 607, A98

\bibitem[{{Duan} {et~al.}(2024){Duan}, {Li}, {Conselice}, {Harvey}, {Austin}, {Adams}, {Ferreira}, {Duncan}, {Trussler}, {Pascalau}, {Windhorst}, {Holwerda}, {Broadhurst}, {Coe}, {Cohen}, {Du}, {Driver}, {Frye}, {Grogin}, {Hathi}, {Jansen}, {Koekemoer}, {Marshall}, {Nonino}, {Ortiz}, {Pirzkal}, {Robotham}, {Ryan}, {Summers}, {D'Silva}, {Willmer}, \& {Yan}}]{2024arXiv241104944D}
{Duan}, Q., {Li}, Q., {Conselice}, C.~J., {et~al.} 2024, arXiv e-prints, arXiv:2411.04944, {MNRAS, submitted}

\bibitem[{{Eckert} {et~al.}(2020){Eckert}, {Finoguenov}, {Ghirardini}, {Grandis}, {Kaefer}, {Sanders}, \& {Ramos-Ceja}}]{2020OJAp....3E..12E}
{Eckert}, D., {Finoguenov}, A., {Ghirardini}, V., {et~al.} 2020, The Open Journal of Astrophysics, 3, 12

\bibitem[{{Fabian} {et~al.}(2023){Fabian}, {Sanders}, {Ferland}, {McNamara}, {Pinto}, \& {Walker}}]{fabian2023}
{Fabian}, A.~C., {Sanders}, J.~S., {Ferland}, G.~J., {et~al.} 2023, \mnras, 524, 716

\bibitem[{{Fabian} {et~al.}(2024){Fabian}, {Sanders}, {Ferland}, {McNamara}, {Pinto}, \& {Walker}}]{fabian2024}
{Fabian}, A.~C., {Sanders}, J.~S., {Ferland}, G.~J., {et~al.} 2024, \mnras, 531, 267

\bibitem[{{Ferr{\'e}-Mateu} {et~al.}(2015){Ferr{\'e}-Mateu}, {Mezcua}, {Trujillo}, {Balcells}, \& {van den Bosch}}]{2015ApJ...808...79F}
{Ferr{\'e}-Mateu}, A., {Mezcua}, M., {Trujillo}, I., {Balcells}, M., \& {van den Bosch}, R. C.~E. 2015, \apj, 808, 79

\bibitem[{{Ferr{\'e}-Mateu} {et~al.}(2017){Ferr{\'e}-Mateu}, {Trujillo}, {Mart{\'\i}n-Navarro}, {Vazdekis}, {Mezcua}, {Balcells}, \& {Dom{\'\i}nguez}}]{2017MNRAS.467.1929F}
{Ferr{\'e}-Mateu}, A., {Trujillo}, I., {Mart{\'\i}n-Navarro}, I., {et~al.} 2017, \mnras, 467, 1929

\bibitem[{{Fruscione} {et~al.}(2006){Fruscione}, {McDowell}, {Allen}, {Brickhouse}, {Burke}, {Davis}, {Durham}, {Elvis}, {Galle}, {Harris}, {Huenemoerder}, {Houck}, {Ishibashi}, {Karovska}, {Nicastro}, {Noble}, {Nowak}, {Primini}, {Siemiginowska}, {Smith}, \& {Wise}}]{2006SPIE.6270E..1VF}
{Fruscione}, A., {McDowell}, J.~C., {Allen}, G.~E., {et~al.} 2006, in Society of Photo-Optical Instrumentation Engineers (SPIE) Conference Series, Vol. 6270, Observatory Operations: Strategies, Processes, and Systems, ed. D.~R. {Silva} \& R.~E. {Doxsey}, 62701V

\bibitem[{{Gabriel}(2017)}]{2017xru..conf...84G}
{Gabriel}, C. 2017, in The X-ray Universe 2017, ed. J.-U. {Ness} \& S.~{Migliari}, 84

\bibitem[{{Gendron-Marsolais} {et~al.}(2017){Gendron-Marsolais}, {Kraft}, {Bogdan}, {Hlavacek-Larrondo}, {Forman}, {Jones}, {Su}, {Nulsen}, {Randall}, \& {Roediger}}]{2017ApJ...848...26G}
{Gendron-Marsolais}, M., {Kraft}, R.~P., {Bogdan}, A., {et~al.} 2017, \apj, 848, 26

\bibitem[{{Gilfanov}(2004)}]{2004MNRAS.349..146G}
{Gilfanov}, M. 2004, \mnras, 349, 146

\bibitem[{{Goudfrooij} \& {de Jong}(1995)}]{1995A&A...298..784G}
{Goudfrooij}, P. \& {de Jong}, T. 1995, \aap, 298, 784

\bibitem[{{Greene} {et~al.}(2024){Greene}, {Labbe}, {Goulding}, {Furtak}, {Chemerynska}, {Kokorev}, {Dayal}, {Volonteri}, {Williams}, {Wang}, {Setton}, {Burgasser}, {Bezanson}, {Atek}, {Brammer}, {Cutler}, {Feldmann}, {Fujimoto}, {Glazebrook}, {de Graaff}, {Khullar}, {Leja}, {Marchesini}, {Maseda}, {Matthee}, {Miller}, {Naidu}, {Nanayakkara}, {Oesch}, {Pan}, {Papovich}, {Price}, {van Dokkum}, {Weaver}, {Whitaker}, \& {Zitrin}}]{2024ApJ...964...39G}
{Greene}, J.~E., {Labbe}, I., {Goulding}, A.~D., {et~al.} 2024, \apj, 964, 39

\bibitem[{{Inayoshi}(2025)}]{2025ApJ...988L..22I}
{Inayoshi}, K. 2025, \apjl, 988, L22

\bibitem[{{Irwin} {et~al.}(2003){Irwin}, {Athey}, \& {Bregman}}]{2003ApJ...587..356I}
{Irwin}, J.~A., {Athey}, A.~E., \& {Bregman}, J.~N. 2003, \apj, 587, 356

\bibitem[{Kaastra {et~al.}(2024)Kaastra, Raassen, de~Plaa, \& Gu}]{kaastra_2024_12771915}
Kaastra, J.~S., Raassen, A. J.~J., de~Plaa, J., \& Gu, L. 2024, SPEX X-ray spectral fitting package

\bibitem[{{Kara} {et~al.}(2024){Kara}, {Pl{\v{s}}ek}, {Protu{\v{s}}ov{\'a}}, {Breuer}, {Werner}, {Mernier}, \& {Ercan}}]{kara2024}
{Kara}, S., {Pl{\v{s}}ek}, T., {Protu{\v{s}}ov{\'a}}, K., {et~al.} 2024, \mnras, 528, 1500

\bibitem[{{Kim} \& {Fabbiano}(2004)}]{2004ApJ...611..846K}
{Kim}, D.-W. \& {Fabbiano}, G. 2004, \apj, 611, 846

\bibitem[{{Kocevski} {et~al.}(2025){Kocevski}, {Finkelstein}, {Barro}, {Taylor}, {Calabr{\`o}}, {Laloux}, {Buchner}, {Trump}, {Leung}, {Yang}, {Dickinson}, {P{\'e}rez-Gonz{\'a}lez}, {Pacucci}, {Inayoshi}, {Somerville}, {McGrath}, {Akins}, {Bagley}, {Bowler}, {Bisigello}, {Carnall}, {Casey}, {Cheng}, {Cleri}, {Costantin}, {Cullen}, {Davis}, {Donnan}, {Dunlop}, {Ellis}, {Ferguson}, {Fujimoto}, {Fontana}, {Giavalisco}, {Grazian}, {Grogin}, {Hathi}, {Hirschmann}, {Huertas-Company}, {Holwerda}, {Illingworth}, {Juneau}, {Kartaltepe}, {Koekemoer}, {Li}, {Lucas}, {Magee}, {Mason}, {McLeod}, {McLure}, {Napolitano}, {Papovich}, {Pirzkal}, {Rodighiero}, {Santini}, {Wilkins}, \& {Yung}}]{2025ApJ...986..126K}
{Kocevski}, D.~D., {Finkelstein}, S.~L., {Barro}, G., {et~al.} 2025, \apj, 986, 126

\bibitem[{{Kov{\'a}cs} {et~al.}(2024){Kov{\'a}cs}, {Bogd{\'a}n}, {Natarajan}, {Werner}, {Azadi}, {Volonteri}, {Tremblay}, {Chadayammuri}, {Forman}, {Jones}, \& {Kraft}}]{2024ApJ...965L..21K}
{Kov{\'a}cs}, O.~E., {Bogd{\'a}n}, {\'A}., {Natarajan}, P., {et~al.} 2024, \apjl, 965, L21

\bibitem[{{Kov{\'a}cs} {et~al.}(2023){Kov{\'a}cs}, {Zhu}, {Werner}, {Simionescu}, \& {Bogd{\'a}n}}]{2023A&A...678A..91K}
{Kov{\'a}cs}, O.~E., {Zhu}, Z., {Werner}, N., {Simionescu}, A., \& {Bogd{\'a}n}, {\'A}. 2023, \aap, 678, A91

\bibitem[{{Labbe} {et~al.}(2025){Labbe}, {Greene}, {Bezanson}, {Fujimoto}, {Furtak}, {Goulding}, {Matthee}, {Naidu}, {Oesch}, {Atek}, {Brammer}, {Chemerynska}, {Coe}, {Cutler}, {Dayal}, {Feldmann}, {Franx}, {Glazebrook}, {Leja}, {Maseda}, {Marchesini}, {Nanayakkara}, {Nelson}, {Pan}, {Papovich}, {Price}, {Suess}, {Wang}, {Weaver}, {Whitaker}, {Williams}, \& {Zitrin}}]{2025ApJ...978...92L}
{Labbe}, I., {Greene}, J.~E., {Bezanson}, R., {et~al.} 2025, \apj, 978, 92

\bibitem[{{Lakhchaura} {et~al.}(2018){Lakhchaura}, {Werner}, {Sun}, {Canning}, {Gaspari}, {Allen}, {Connor}, {Donahue}, \& {Sarazin}}]{2018MNRAS.481.4472L}
{Lakhchaura}, K., {Werner}, N., {Sun}, M., {et~al.} 2018, \mnras, 481, 4472

\bibitem[{{Libeskind} {et~al.}(2011){Libeskind}, {Knebe}, {Hoffman}, {Gottl{\"o}ber}, {Yepes}, \& {Steinmetz}}]{2011MNRAS.411.1525L}
{Libeskind}, N.~I., {Knebe}, A., {Hoffman}, Y., {et~al.} 2011, \mnras, 411, 1525

\bibitem[{Lin {et~al.}(2024)Lin, Zheng, Jiang, Yuan, Ho, Wang, Jiang, Rhoads, Malhotra, Barrientos, Wold, Infante, Zhu, Ji, \& Fu}]{lin2024discoverylocalanalogsjwsts}
Lin, R., Zheng, Z.-Y., Jiang, C., {et~al.} 2024, Discovery of Local Analogs to JWST's Little Red Dots

\bibitem[{{Lisiecki} {et~al.}(2023){Lisiecki}, {Ma{\l}ek}, {Siudek}, {Pollo}, {Krywult}, {Karska}, \& {Junais}}]{2023A&A...669A..95L}
{Lisiecki}, K., {Ma{\l}ek}, K., {Siudek}, M., {et~al.} 2023, \aap, 669, A95

\bibitem[{{Liu} {et~al.}(2022){Liu}, {Bulbul}, {Ghirardini}, {Liu}, {Klein}, {Clerc}, {{\"O}zsoy}, {Ramos-Ceja}, {Pacaud}, {Comparat}, {Okabe}, {Bahar}, {Biffi}, {Brunner}, {Br{\"u}ggen}, {Buchner}, {Ider Chitham}, {Chiu}, {Dolag}, {Gatuzz}, {Gonzalez}, {Hoang}, {Lamer}, {Merloni}, {Nandra}, {Oguri}, {Ota}, {Predehl}, {Reiprich}, {Salvato}, {Schrabback}, {Sanders}, {Seppi}, \& {Thibaud}}]{2022A&A...661A...2L}
{Liu}, A., {Bulbul}, E., {Ghirardini}, V., {et~al.} 2022, \aap, 661, A2

\bibitem[{{Lodders} \& {Palme}(2009)}]{lodders2009}
{Lodders}, K. \& {Palme}, H. 2009, Meteoritics and Planetary Science Supplement, 72, 5154

\bibitem[{{Maiolino} {et~al.}(2024){Maiolino}, {Scholtz}, {Curtis-Lake}, {Carniani}, {Baker}, {de Graaff}, {Tacchella}, {{\"U}bler}, {D'Eugenio}, {Witstok}, {Curti}, {Arribas}, {Bunker}, {Charlot}, {Chevallard}, {Eisenstein}, {Egami}, {Ji}, {Jones}, {Lyu}, {Rawle}, {Robertson}, {Rujopakarn}, {Perna}, {Sun}, {Venturi}, {Williams}, \& {Willott}}]{2024A&A...691A.145M}
{Maiolino}, R., {Scholtz}, J., {Curtis-Lake}, E., {et~al.} 2024, \aap, 691, A145

\bibitem[{{Mathews} \& {Brighenti}(2003)}]{2003ARA&A..41..191M}
{Mathews}, W.~G. \& {Brighenti}, F. 2003, \araa, 41, 191

\bibitem[{{Mathews} {et~al.}(2006){Mathews}, {Brighenti}, {Faltenbacher}, {Buote}, {Humphrey}, {Gastaldello}, \& {Zappacosta}}]{2006ApJ...652L..17M}
{Mathews}, W.~G., {Brighenti}, F., {Faltenbacher}, A., {et~al.} 2006, \apjl, 652, L17

\bibitem[{{Matthee} {et~al.}(2024){Matthee}, {Naidu}, {Brammer}, {Chisholm}, {Eilers}, {Goulding}, {Greene}, {Kashino}, {Labbe}, {Lilly}, {Mackenzie}, {Oesch}, {Weibel}, {Wuyts}, {Xiao}, {Bordoloi}, {Bouwens}, {van Dokkum}, {Illingworth}, {Kramarenko}, {Maseda}, {Mason}, {Meyer}, {Nelson}, {Reddy}, {Shivaei}, {Simcoe}, \& {Yue}}]{matthee2024}
{Matthee}, J., {Naidu}, R.~P., {Brammer}, G., {et~al.} 2024, \apj, 963, 129

\bibitem[{{Mernier} {et~al.}(2018){Mernier}, {Werner}, {de Plaa}, {Kaastra}, {Raassen}, {Gu}, {Mao}, {Urdampilleta}, \& {Simionescu}}]{mernier2018}
{Mernier}, F., {Werner}, N., {de Plaa}, J., {et~al.} 2018, \mnras, 480, L95

\bibitem[{{Naab} {et~al.}(2009){Naab}, {Johansson}, \& {Ostriker}}]{2009ApJ...699L.178N}
{Naab}, T., {Johansson}, P.~H., \& {Ostriker}, J.~P. 2009, \apjl, 699, L178

\bibitem[{{Oser} {et~al.}(2010){Oser}, {Ostriker}, {Naab}, {Johansson}, \& {Burkert}}]{2010ApJ...725.2312O}
{Oser}, L., {Ostriker}, J.~P., {Naab}, T., {Johansson}, P.~H., \& {Burkert}, A. 2010, \apj, 725, 2312

\bibitem[{{O'Sullivan} {et~al.}(2003){O'Sullivan}, {Ponman}, \& {Collins}}]{2003MNRAS.340.1375O}
{O'Sullivan}, E., {Ponman}, T.~J., \& {Collins}, R.~S. 2003, \mnras, 340, 1375

\bibitem[{{Park} {et~al.}(2006){Park}, {Kashyap}, {Siemiginowska}, {van Dyk}, {Zezas}, {Heinke}, \& {Wargelin}}]{2006ApJ...652..610P}
{Park}, T., {Kashyap}, V.~L., {Siemiginowska}, A., {et~al.} 2006, \apj, 652, 610

\bibitem[{{Peralta de Arriba} {et~al.}(2016){Peralta de Arriba}, {Quilis}, {Trujillo}, {Cebri{\'a}n}, \& {Balcells}}]{2016MNRAS.461..156P}
{Peralta de Arriba}, L., {Quilis}, V., {Trujillo}, I., {Cebri{\'a}n}, M., \& {Balcells}, M. 2016, \mnras, 461, 156

\bibitem[{{P{\'e}rez-Gonz{\'a}lez} {et~al.}(2024){P{\'e}rez-Gonz{\'a}lez}, {Barro}, {Rieke}, {Lyu}, {Rieke}, {Alberts}, {Williams}, {Hainline}, {Sun}, {Pusk{\'a}s}, {Annunziatella}, {Baker}, {Bunker}, {Egami}, {Ji}, {Johnson}, {Robertson}, {Rodr{\'\i}guez Del Pino}, {Rujopakarn}, {Shivaei}, {Tacchella}, {Willmer}, \& {Willott}}]{2024ApJ...968....4P}
{P{\'e}rez-Gonz{\'a}lez}, P.~G., {Barro}, G., {Rieke}, G.~H., {et~al.} 2024, \apj, 968, 4

\bibitem[{{Quilis} \& {Trujillo}(2013)}]{2013ApJ...773L...8Q}
{Quilis}, V. \& {Trujillo}, I. 2013, \apjl, 773, L8

\bibitem[{{Rantala} {et~al.}(2024){Rantala}, {Rawlings}, {Naab}, {Thomas}, \& {Johansson}}]{2024MNRAS.535.1202R}
{Rantala}, A., {Rawlings}, A., {Naab}, T., {Thomas}, J., \& {Johansson}, P.~H. 2024, \mnras, 535, 1202

\bibitem[{{Reines} \& {Volonteri}(2015)}]{2015ApJ...813...82R}
{Reines}, A.~E. \& {Volonteri}, M. 2015, \apj, 813, 82

\bibitem[{{Shen} {et~al.}(2003){Shen}, {Mo}, {White}, {Blanton}, {Kauffmann}, {Voges}, {Brinkmann}, \& {Csabai}}]{2003MNRAS.343..978S}
{Shen}, S., {Mo}, H.~J., {White}, S. D.~M., {et~al.} 2003, \mnras, 343, 978

\bibitem[{{Svoboda} {et~al.}(2019){Svoboda}, {Douna}, {Orlitov{\'a}}, \& {Ehle}}]{2019ApJ...880..144S}
{Svoboda}, J., {Douna}, V., {Orlitov{\'a}}, I., \& {Ehle}, M. 2019, \apj, 880, 144

\bibitem[{{Trujillo} {et~al.}(2014){Trujillo}, {Ferr{\'e}-Mateu}, {Balcells}, {Vazdekis}, \& {S{\'a}nchez-Bl{\'a}zquez}}]{2014ApJ...780L..20T}
{Trujillo}, I., {Ferr{\'e}-Mateu}, A., {Balcells}, M., {Vazdekis}, A., \& {S{\'a}nchez-Bl{\'a}zquez}, P. 2014, \apjl, 780, L20

\bibitem[{{Voit} {et~al.}(2005){Voit}, {Kay}, \& {Bryan}}]{2005MNRAS.364..909V}
{Voit}, G.~M., {Kay}, S.~T., \& {Bryan}, G.~L. 2005, \mnras, 364, 909

\bibitem[{{Wellons} {et~al.}(2015){Wellons}, {Torrey}, {Ma}, {Rodriguez-Gomez}, {Vogelsberger}, {Kriek}, {van Dokkum}, {Nelson}, {Genel}, {Pillepich}, {Springel}, {Sijacki}, {Snyder}, {Nelson}, {Sales}, \& {Hernquist}}]{2015MNRAS.449..361W}
{Wellons}, S., {Torrey}, P., {Ma}, C.-P., {et~al.} 2015, \mnras, 449, 361

\bibitem[{{Werner} {et~al.}(2012){Werner}, {Allen}, \& {Simionescu}}]{2012MNRAS.425.2731W}
{Werner}, N., {Allen}, S.~W., \& {Simionescu}, A. 2012, \mnras, 425, 2731

\bibitem[{{Werner} {et~al.}(2018){Werner}, {Lakhchaura}, {Canning}, {Gaspari}, \& {Simionescu}}]{2018MNRAS.477.3886W}
{Werner}, N., {Lakhchaura}, K., {Canning}, R.~E.~A., {Gaspari}, M., \& {Simionescu}, A. 2018, \mnras, 477, 3886

\bibitem[{{Werner} {et~al.}(2019){Werner}, {McNamara}, {Churazov}, \& {Scannapieco}}]{2019SSRv..215....5W}
{Werner}, N., {McNamara}, B.~R., {Churazov}, E., \& {Scannapieco}, E. 2019, \ssr, 215, 5

\bibitem[{{Willingale} {et~al.}(2013){Willingale}, {Starling}, {Beardmore}, {Tanvir}, \& {O'Brien}}]{2013MNRAS.431..394W}
{Willingale}, R., {Starling}, R.~L.~C., {Beardmore}, A.~P., {Tanvir}, N.~R., \& {O'Brien}, P.~T. 2013, \mnras, 431, 394

\bibitem[{{Y{\i}ld{\i}r{\i}m} {et~al.}(2017){Y{\i}ld{\i}r{\i}m}, {van den Bosch}, {van de Ven}, {Mart{\'\i}n-Navarro}, {Walsh}, {Husemann}, {G{\"u}ltekin}, \& {Gebhardt}}]{2017MNRAS.468.4216Y}
{Y{\i}ld{\i}r{\i}m}, A., {van den Bosch}, R. C.~E., {van de Ven}, G., {et~al.} 2017, \mnras, 468, 4216

\bibitem[{{Zhang} {et~al.}(2024){Zhang}, {Comparat}, {Ponti}, {Merloni}, {Nandra}, {Haberl}, {Truong}, {Pillepich}, {Locatelli}, {Zhang}, {Sanders}, {Zheng}, {Liu}, {Popesso}, {Liu}, {Predehl}, {Salvato}, {Shreeram}, \& {Yeung}}]{2024A&A...690A.268Z}
{Zhang}, Y., {Comparat}, J., {Ponti}, G., {et~al.} 2024, \aap, 690, A268

\bibitem[{{Zhang} {et~al.}(2012){Zhang}, {Gilfanov}, \& {Bogd{\'a}n}}]{2012A&A...546A..36Z}
{Zhang}, Z., {Gilfanov}, M., \& {Bogd{\'a}n}, {\'A}. 2012, \aap, 546, A36

\bibitem[{{Zhu} {et~al.}(2023){Zhu}, {Kov{\'a}cs}, {Simionescu}, \& {Werner}}]{2023A&A...678A.122Z}
{Zhu}, Z., {Kov{\'a}cs}, O.~E., {Simionescu}, A., \& {Werner}, N. 2023, \aap, 678, A122

\end{thebibliography}

\end{document}